\documentclass{aastex701}
\usepackage{longtable}
\usepackage{makecell}
\usepackage{hyperref}
\begin{document}

\title{Statistical Study of Solar Prominence Plumes Based on NVST H$\alpha$ Observations}

\correspondingauthor{Yijun Hou}
\email{yijunhou@nao.cas.cn}

\author{Yangrui Chen}
\affiliation{School of Physical Science and Technology, Southwest Jiaotong University, Chengdu 611756, China}
\affiliation{State Key Laboratory of Solar Activity and Space Weather, National Astronomical Observatories, Chinese Academy of Science, Beijing 100101, China}
\email{chenyangrui@my.swjtu.edu.cn}

\author[0000-0002-9534-1638]{Yijun Hou}
\affiliation{State Key Laboratory of Solar Activity and Space Weather, National Astronomical Observatories, Chinese Academy of Science, Beijing 100101, China}
\affiliation{School of Astronomy and Space Science, University of Chinese Academy of Sciences, Beijing 100049, China}
\email{yijunhou@nao.cas.cn}

\author{Jincheng Wang}
\affiliation{Yunnan Observatories, Chinese Academy of Sciences, Kunming Yunnan 650216, China}
\affiliation{Yunnan Key Laboratory of Solar Physics and Space Science, Kunming 650216, China}
\email{wangjincheng@ynao.ac.cn} 

\author[0000-0001-6655-1743]{Ting Li}
\affiliation{State Key Laboratory of Solar Activity and Space Weather, National Space Science Center, Chinese Academy of Sciences, Beijing 100190, China}
\affiliation{School of Astronomy and Space Science, University of Chinese Academy of Sciences, Beijing 100049, China}
\email{liting@nssc.ac.cn}

\author{Shuo Yang}
\affiliation{North China University of Technology, Beijing 100144, China}
\email{002459@ncut.edu.cn} 

\author[0000-0001-9893-1281]{Yilin Guo}
\affiliation{Beijing Planetarium, Beijing Academy of Science and Technology, Beijing 100044, China}
\email{guoyilin@bjp.org.cn}

\author[0009-0004-4079-1342]{Junyi Zhang}
\affiliation{State Key Laboratory of Solar Activity and Space Weather, National Astronomical Observatories, Chinese Academy of Science, Beijing 100101, China}
\affiliation{School of Astronomy and Space Science, University of Chinese Academy of Sciences, Beijing 100049, China}
\email{zhangjunyi@nao.cas.cn}

\author[0000-0001-6024-8399]{Haitang Li}
\affiliation{School of Physical Science and Technology, Southwest Jiaotong University, Chengdu 611756, China}
\email{lihaitang@swjtu.edu.cn}

\author[0000-0002-2995-070X]{Feiyang Sha}
\affiliation{School of Physical Science and Technology, Southwest Jiaotong University, Chengdu 611756, China}
\email{shafy@my.swjtu.edu.cn}  

\author{Qing Zhou}
\affiliation{School of Physical Science and Technology, Southwest Jiaotong University, Chengdu 611756, China}
\email{zhouqing@my.swjtu.edu.cn}
	
\author[0000-0002-7694-2454]{Yu Liu}
\affiliation{School of Physical Science and Technology, Southwest Jiaotong University, Chengdu 611756, China}
\email{lyu@swjtu.edu.cn}

\author{Xiaoli Yan}
\affiliation{Yunnan Observatories, Chinese Academy of Sciences, Kunming Yunnan 650216, China}
\affiliation{Yunnan Key Laboratory of Solar Physics and Space Science, Kunming 650216, China}
\email{yanxl@ynao.ac.cn}

\begin{abstract}
Plumes are one of the most representative dynamic features observed in prominences and play a key role in mass and magnetic transport within them. However, their physical nature and triggering processes remain actively debated. Based on limb H$\alpha$ observations from the New Vacuum Solar Telescope (NVST) during 2013--2025, we statistically investigated 34 plumes with clear and complete evolutions by developing an automated image-processing pipeline. It is revealed that plume lifetimes mainly range from 300 s to 700 s, with vertical displacements between 3--7 Mm. The mean widths and velocities are concentrated in the range of 0.5--1.5 Mm and 10--20 km s$^{-1}$, respectively. Besides wide distribution ranges, plume parameters exhibit irregular evolution fluctuations, indicating that the formation and evolution of various plumes may exhibit different physical patterns. Correlation analysis among the parameters further reveals that: (1) Positive correlations were found among lifetime, vertical displacement, and mean width, indicating an intrinsic coupling between the temporal and spatial scales of plumes. (2) Trajectory curvature is negatively correlated with lifetime, vertical displacement, and velocity. Accelerating and width-contracting plumes typically have lower curvature, suggesting that curvature may reflect environmental influences and the stability of plumes. (3) Plumes with higher initial velocities were more likely to be accompanied by precursor brightening, suggesting that these plumes may be triggered by magnetic reconnection. Furthermore, we infer that some plumes in non-bubble regions may be inherently driven by mini-filament eruptions. These results establish a statistical framework for prominence plumes and reveal diversity in their dynamical evolution and triggering mechanisms.
\end{abstract}
\keywords{Solar activity (1475) --- Solar prominences (1519) --- Solar magnetic fields (1503)}

\section{Introduction} 

Solar prominences are composed of relatively cool and dense plasma embedded in the hot and rarefied solar corona. In H$\alpha$ observations, they are seen as bright structures extending above the solar limb, whereas on the solar disk they appear as dark features against the bright background and are commonly termed filaments. They are generally thought to be supported by highly sheared or twisted magnetic field configurations above magnetic polarity inversion lines (PILs) \citep{2010SSRv..151..243L, 2020RAA....20..166C, 2023ApJ...959...69H}. Extensive observations have demonstrated that prominences are closely associated with eruptive phenomena such as solar flares and coronal mass ejections (CMEs), serving as important carriers for the accumulation and release of magnetic energy in the solar atmosphere \citep{2015SoPh..290.1703M, 2020A&A...640A.101H, 2025ApJ...993..126L, 2026ApJ..1004...66Z}.

In recent years, with the advancement of high-resolution solar telescopes, a variety of dynamical phenomena within solar prominences have been resolved, including horizontal, vertical, and counterstreaming plasma flows \citep{2008ApJ...676L..89B,2015ApJ...814L..17S,2025ApJ...987...38Z}, transverse swaying motions \citep{2014ApJ...795..130S,2018ApJ...863..192L,2024ApJ...965L..28W}, bubbles \citep{2011Natur.472..197B,2021ApJ...911L...9G,2024ApJ...970..110G}, and rising plumes \citep{2008ApJ...676L..89B,2010ApJ...716.1288B,2025ApJ...990L..64W}. Among these, bubbles and plumes are the most representative dynamical features observed in the basal boundary region of prominences. They not only reflect the instability of the local magnetic field and plasma in the prominence boundary layer, but may also influence the stability of the prominence through processes such as mass transport, magnetic topology evolution, and localized magnetic energy release \citep{2008ApJ...676L..89B,2010ApJ...716.1288B,2012ApJ...761....9D}. Prominence bubbles appear as low-emission, cavity-like structures in H$\alpha$ and EUV observations. Differential emission measure (DEM) analyzes indicate that bubbles are generally hotter than the surrounding prominence plasma and exhibit temperatures closer to those of the ambient corona \citep{2015ApJ...814L..17S}. One view suggests that the interior of a bubble is filled with relatively hot, low-density plasma \citep{2011Natur.472..197B}, while another view proposes that it is more akin to a plasma cavity \citep{2012ApJ...761....9D}. 

Regarding the formation mechanism of prominence bubbles, the prevailing interpretation is that they are associated with magnetic flux emergence beneath them \citep{2011Natur.472..197B,2015ApJ...814L..17S}. Magnetic topology analysis further indicates that the inclusion of parasitic bipoles in the background magnetic field can form arch-shaped bubble structures consistent with observations \citep{2012ApJ...761....9D,2014A&A...567A.123G}. Combining multi-perspective observations of an on-disk bubble with nonlinear force-free field (NLFFF) extrapolations, \cite{2021ApJ...911L...9G} showed that the bubble boundary corresponds to an arch-shaped interface with underlying magnetic loops rooted in a surrounding photospheric magnetic patch. More recently, three-dimensional magneto-frictional simulations suggest that prominence bubbles may form spontaneously during the evolution of a magnetic flux rope driven by supergranular motions, without requiring parasitic polarities or embedded small bipoles \citep{2026SCPMA..6949611C}. Besides the traditionally observed quasi-stable bubbles (Type I), \cite{2024ApJ...970..110G} proposed that there is another new type of bubbles, termed transient bubbles (Type II). Type-II bubbles are characterized by shorter lifetimes and faster expansion speeds, and are triggered by the eruption of mini-filaments, accompanied by rapid evacuation of local plasma. When the angle between the axis of erupting mini-filament and the line of sight (LOS) is sufficiently small, such transient bubbles may also manifest observationally as plume-like structures.

Prominence plumes are dark, upward-moving structures that usually originate near bubble boundaries. They are commonly considered to be manifestations of plasma instabilities and may serve as important channels for upward plasma transport within prominences \citep{2010ApJ...716.1288B,2014LRSP...11....1P}. However, their physical origin remains under debate. The prevailing view suggests that plumes are driven by plasma instabilities \citep{2010SoPh..267...75R,2010ApJ...716.1288B,2018RvMPP...2....1H}. Magnetohydrodynamics (MHD) numerical studies introduced a density inversion at the prominence–cavity interface to investigate the nonlinear evolution of these instabilities, showing that the magnetic Rayleigh-Taylor (RT) instability could generate upward-moving, finger-like, or mushroom-shaped structures. Their ascent, fragmentation, and multiscale fine structures were broadly consistent with observed plume morphologies \citep{2012ApJ...746..120H,2012ApJ...756..110H}. However, RT instability alone cannot fully explain the vortical motions and complex morphologies observed at plume fronts. To address this, some studies have introduced the coupling effect of Kelvin–Helmholtz (KH) instability and RT instability under velocity shear conditions to explain the curling and mixing process at the interface \citep{2010SoPh..267...75R,2017ApJ...850...60B}. Meanwhile, recent observations suggested that enhanced spicule activity beneath prominences could drive shock waves and trigger magnetic Richtmyer-Meshkov instability (RMI) at the prominence-bubble interface, thereby leading to the formation of small-scale plumes \citep{2025ApJ...990L..64W}.

In contrast to the framework dominated by MHD instabilities, another class of model suggested that the role of rising mini-filaments or magnetic flux ropes was a key driver of plume formation. Some studies suggested that magnetic reconnection occurred between erupting mini-filaments/flux ropes and the prominence magnetic fields at the bubble boundaries, thereby triggering plumes at these boundaries \citep{2012ApJ...761....9D,2014A&A...567A.123G,2015ApJ...814L..17S,2022A&A...659A..76W}. Alternatively, it was proposed that the interaction between erupting mini-filaments/flux ropes and the prominence magnetic fields could also drive plume formation in the absence of magnetic reconnection \citep{2021ApJ...923L..10C,2021RAA....21..222X}. In addition, \cite{2024ApJ...970..110G} proposed that transient bubbles (Type-II), triggered by mini-filament eruptions, might also appear as plume structures under specific observational perspectives during their rapid expansion.

Although significant progress has been made in recent years in the study of solar prominence plumes, most existing observational works are still limited to individual case studies or rare small-sample analyzes. Owing to differences in observing conditions, spatial resolution, and analysis techniques, the reported physical properties of plumes exhibit substantial variability, resulting in a lack of consensus regarding their physical nature. Moreover, most previous studies have concentrated on plume formation or localized dynamical evolution, whereas statistical investigations of the relationships among plume morphology, kinematic properties, and triggering environments remain relatively scarce. Such studies, however, are essential for understanding mass and energy transport within prominences and their implications for prominence stability.

To address these issues, we constructed a prominence plume catalog using high-quality limb-prominence observations obtained by the New Vacuum Solar Telescope (NVST; \citealp{2014RAA....14..705L,2020ScChE..63.1656Y}) between 2013 and 2025. Based on this catalog, we perform a statistical investigation of the morphological and dynamical properties of prominence plumes, aiming to characterize their evolutionary behavior and explore their possible triggering mechanisms. The remainder of this paper is organized as follows. Section 2 describes the observational data, sample selection criteria, and data reduction procedures. Section 3 presents the statistical results of the plume parameters and discusses their evolutionary characteristics and relationships with the triggering environment. Finally, Section 4 summarizes the main conclusions and discusses their implications.

\section{Observations and Methods} 
\subsection{NVST Observations of  Prominences}
NVST is a ground-based solar telescope with high spatial resolution. Its H$\alpha$ channel is equipped with a tunable Lyot filter with a bandwidth of approximately 0.25 {\AA}, enabling spectral scanning observations and thus allowing detailed investigations of the formation and evolution of solar filaments, prominences, and their fine structures. The reconstructed H$\alpha$ images achieve high spatial and temporal resolutions, making NVST suited for studies of the fine structures and dynamical behaviors of solar prominences \citep{2014RAA....14..705L}. Based on the high-quality limb-prominence observations obtained by NVST, numerous studies on prominence fine structures and their dynamical evolution have been conducted in recent years, including investigations of prominence oscillations, thread flows, bubbles, and plumes \citep{2015ApJ...814L..17S,2018ApJ...863..192L,2019ApJ...884L..51Y,2021ApJ...911L...9G,2022A&A...659A..76W,2024ApJ...970..110G,2024ApJ...965L..28W,2025ApJ...981..139Y}.

\subsection{Construction of the Plume catalog}
Based on all limb prominence observations obtained by NVST during 2013--2025, we constructed a catalog containing 201 prominence samples. Using this catalog, all prominences were carefully inspected manually, and events with clearly identifiable plume formation and subsequent upward evolution were selected, yielding a total of 36 prominences. Within these 36 prominences, 137 plumes were identified, as multiple plume events may occur successively or simultaneously within a single prominence and may originate from different source regions, these plumes were treated as independent samples, from which a catalog of plume events was established. According to their locations of origin, plumes were classified into two categories: 59.9\% originated at bubble boundaries, while the remaining 40.1\% formed in regions not associated with bubble boundaries. Both the catalog of prominence samples and the catalog of plume events have been compiled as an integrated dataset and deposited in the Science Data Bank (ScienceDB) repository (\url{https://doi.org/10.57760/sciencedb.39712}).

To perform a detailed statistical analysis of plume evolution, the plume samples were further filtered according to the following criteria:
\begin{enumerate}
\item The plume exhibited a complete and clearly defined boundary during its formation stage;
\item The plume was not significantly obscured along the LOS by surrounding prominence threads or other structures;
\item The plume remained an independent structure throughout its evolution and did not merge with pre-existing gaps between prominence threads or with other plumes;
\item The entire formation process, from the initial appearance of the plume to its subsequent development, was continuously observed without significant interruptions in the time series.
\end{enumerate}
Under these constraints, 34 plume events with clear and complete morphological evolution were finally selected for subsequent extraction and statistical analysis of physical parameters.

In H$\alpha$ images, plumes generally appear as elongated cavity-like structures embedded within the brighter background prominence. In this study, the plume boundary was defined as the location of the maximum local radiation intensity gradient. To accurately extract the plume boundary and derive related physical parameters, we developed a complete image-processing and parameter-extraction pipeline as follows:

\begin{enumerate}
\item \textbf{Plume boundary extraction.} The images were first normalized and denoised using the BM3D algorithm \citep{2007ITIP...16.2080D}. The plume boundaries were then enhanced using the Difference of Gaussian (DoG) method \citep{1980RSPSB.207..187M}, Since the DoG-processed data are insensitive to moderate variations in the binarization threshold, a fixed threshold was adopted to identify all candidate plume structures. Morphological operations and connected-component analysis were subsequently applied to suppress noise and isolate the largest connected region, which was identified as the plume region. The initial plume boundary was extracted from this region using a contour-detection algorithm. The extracted boundary was morphologically dilated into a finite‑width band to reduce uncertainties associated with the one‑pixel‑wide contour and to provide a more robust region for subsequent intensity measurements. This finite‑width boundary band was then used to derive boundary coordinates for quantitative analysis.

\item \textbf{Plume boundary segmentation.} The front region of the plume was defined as the set of contour points located near the plume apex, specifically those within two pixels below the maximum height. The left flank and right flank regions were obtained by vertically dividing the contour at the apex and excluding points belonging to the front region.

\item \textbf{Parameter extraction.} For each plume event, the first frame was defined as the earliest frame from which the plume front structure could be stably identified, and the front height exhibited a strictly increasing trend over at least three consecutive frames. The last frame was defined as the latest frame in which the plume front and boundary remained identifiable. The plume lifetime ($\mathrm{Lifetime}$) was defined as the time difference between the last and first frames. The H$\alpha$ images were first rotated such that the local solar limb was aligned horizontally, ensuring that the radial direction pointed upward. Then the vertical displacement ($\mathrm{Height}$) was then defined as the difference between the maximum plume heights measured in the first and last frames along this radial direction. Based on the extracted point sets for each region, the corresponding radiation intensities were sampled from the original images and averaged to obtain the mean radiation intensity. Repeating this procedure for all frames yielded the time series of the plume-front region, left flank region, and right flank region radiation intensities ($\mathrm{I}_{\rm front}$, $\mathrm{I}_{\rm left}$, and $\mathrm{I}_{\rm right}$), respectively. For each frame, the plume width was measured at every pixel height within the plume region. At a given pixel height, the width was defined as the horizontal distance between the leftmost and rightmost pixels belonging to the plume region. The average of all width measurements within that frame was taken as the frame mean width, whereas the largest value was defined as the frame maximum width. Repeating this procedure for all frames produced time series of plume mean width ($\mathrm{Width}_{\rm mean}(t)$) and maximum width ($\mathrm{Width}_{\rm max}(t)$). The event mean width ($\mathrm{Width}_{\rm mean}$) was defined as the average of the frame mean widths over the plume lifetime, while the event maximum width ($\mathrm{Width}_{\rm max}$) was taken as the largest width measured during the entire evolution. The mean widths in the first frame were defined as the initial width ($\mathrm{Width}_{\rm ini}$). The temporal evolution of the centroid of the front region was used to derive the plume trajectory in image coordinates. The velocity time series ($\mathrm{Velocity}(t)$) was constructed from the displacement of the front centroid between consecutive frames divided by the corresponding time interval. The mean velocity ($\mathrm{Velocity}_{\rm mean}$) was defined as the total displacement over the plume lifetime divided by the total time interval. The initial velocity ($\mathrm{Velocity}_{\rm ini}$) was defined as the displacement between the first two frames divided by the corresponding time interval. The maximum velocity ($\mathrm{Velocity}_{\rm max}$) was defined as the maximum value of the velocity time series during the plume evolution. To quantitatively characterize the degree of trajectory curvature, all trajectories were first uniformly smoothed using a Savitzky-Golay filter (window length = 7, polynomial order = 2) to reduce local abrupt fluctuations caused by image noise and limited spatial resolution. The mean curvature ($\mathrm{Curvature}$) was then calculated from the smoothed trajectories using the parametric-curve curvature formula.
\end{enumerate}

\begin{figure*}[h]
\centering
\includegraphics[width=1.\textwidth]{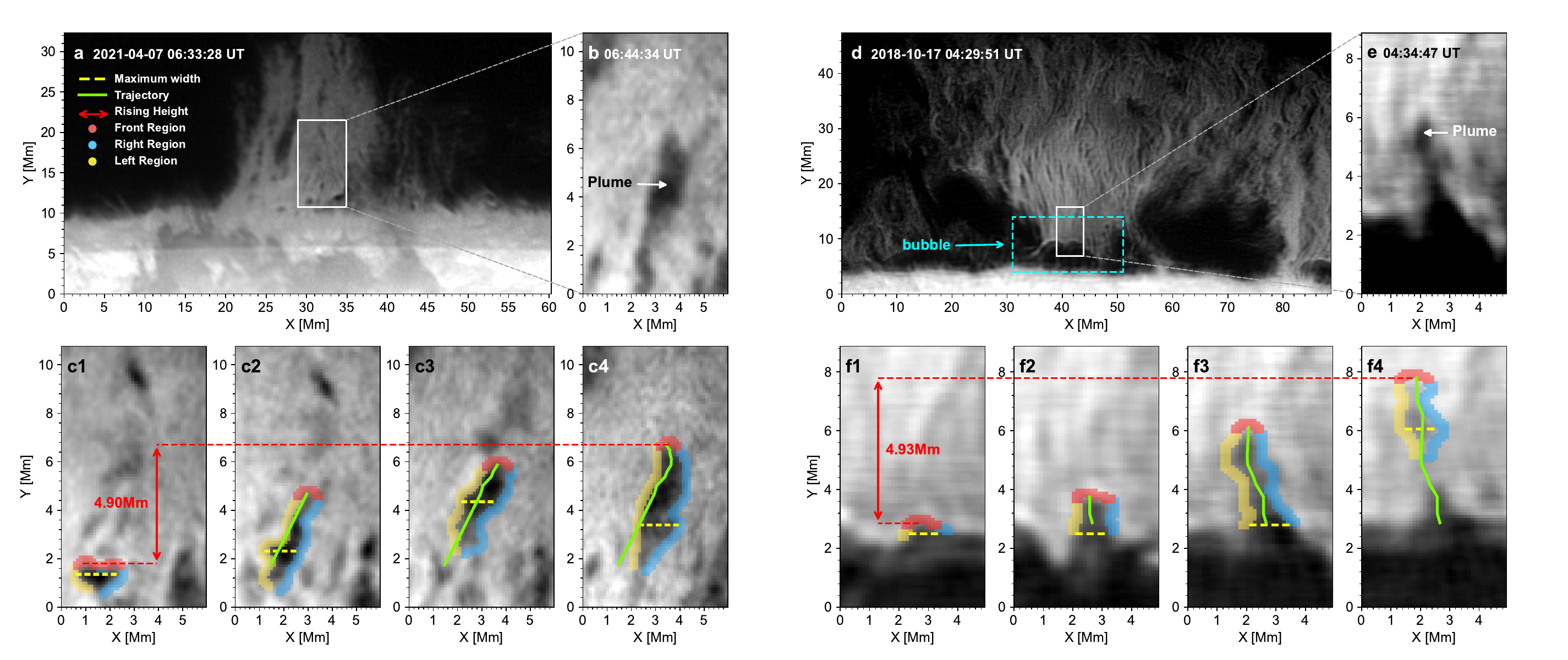}
\caption{Identification results of two representative plumes. Left panels (a–c4) shows a non-bubble plume, and Right panels (d–f4) shows a bubble plume. (a, d) Large-scale context views. The white boxes outline the plume regions, while the cyan dashed box in (d) indicates the underlying bubble. (b, e) Zoomed-in views of the plumes (as indicated by white arrows). (c1–c4, f1–f4) Temporal evolution tracking of the plumes. Colors represent the plume front (red), left flank (yellow), and right flank (blue). Green curves show the trajectories, and yellow dashed lines mark the maximum plume width. Red double-headed arrows indicate the vertical displacement.}
\label{fig:Figure1}
\end{figure*}

To illustrate the parameter-extraction procedure, Figure~\ref{fig:Figure1} presents two representative examples of plumes originating from different locations: a bubble plume and a non-bubble plume. The figure displays the extracted plume boundary contours, motion trajectories, vertical displacements, and the locations of maximum plume width. Using this method, all 34 plume events were processed in a uniform manner, and their morphological and dynamical parameters were extracted. The results are summarized in Table~\ref{tab:plume}.

\begin{deluxetable*}{cccccccccccc}[h]
\tabletypesize{\scriptsize}
\tablewidth{0pt}
\tablecaption{Physical properties of prominence plumes}
\label{tab:plume}
\tablehead{
\colhead{Case} &
\colhead{Time Range} &
\colhead{\makecell{$\mathrm{Lifetime}$\\(s)}} &
\colhead{\makecell{Height\\(Mm)}} &
\colhead{\makecell{$\mathrm{Velocity}_{\rm ini}$\\(km\,s$^{-1}$)}} &
\colhead{\makecell{$\mathrm{Velocity}_{\rm mean}$\\(km\,s$^{-1}$)}} &
\colhead{\makecell{$\mathrm{Velocity}_{\rm max}$\\(km\,s$^{-1}$)}} &
\colhead{\makecell{$\mathrm{Width}_{\rm ini}$\\(Mm)}} &
\colhead{\makecell{$\mathrm{Width}_{\rm mean}$\\(Mm)}} &
\colhead{\makecell{$\mathrm{Width}_{\rm max}$\\(Mm)}} &
\colhead{\makecell{Loc.}} &
\colhead{Curv.}
}
\startdata
1  & \makecell{2016-11-11 \\ 03:18--03:23} & 318.00  & 5.14  & 21.28 & 21.73 & 53.72 & 2.82 & 1.52 & 5.02 & N & C \\
2  & \makecell{2016-11-11 \\ 02:39--02:43} & 242.00  & 4.07  & 27.83 & 20.66 & 55.45 & 0.74 & 0.69 & 1.20 & N & C \\
3  & \makecell{2016-11-11 \\ 02:53--02:58} & 329.00  & 3.59  & 48.63 & 17.98 & 48.63 & 0.56 & 0.69 & 1.20 & N & C \\
4  & \makecell{2018-10-17 \\ 04:30--04:36} & 323.00  & 4.93  & 10.75 & 15.83 & 29.53 & 0.64 & 0.68 & 1.28 & B & S \\
5  & \makecell{2018-10-17 \\ 05:21--05:27} & 323.00  & 4.24  & 65.20 & 17.82 & 65.20 & 2.57 & 1.67 & 5.52 & B & C \\
6  & \makecell{2018-03-19 \\ 07:25--07:31} & 347.87  & 4.14  & 20.85 & 15.25 & 50.87 & 0.63 & 0.53 & 0.89 & N & C \\
7  & \makecell{2018-03-19 \\ 07:46--07:51} & 303.43  & 4.63  & 12.24 & 18.25 & 35.40 & 0.67 & 0.76 & 1.38 & N & C \\
8  & \makecell{2018-11-10 \\ 04:52--04:57} & 327.96  & 4.07  & 20.57 & 14.72 & 20.57 & 0.97 & 1.04 & 2.75 & B & S \\
9  & \makecell{2018-11-10 \\ 05:03--05:14} & 652.87  & 9.09  & 12.75 & 15.81 & 33.23 & 3.28 & 2.33 & 6.34 & B & S \\
10 & \makecell{2018-11-10 \\ 06:45--06:51} & 323.00  & 5.62  & 27.11 & 19.85 & 30.43 & 0.87 & 0.96 & 1.79 & B & S \\
11 & \makecell{2018-11-29 \\ 06:32--06:45} & 798.15  & 8.68  & 3.24  & 13.57 & 35.79 & 4.02 & 4.23 & 6.51 & N & S \\
12 & \makecell{2019-10-28 \\ 07:20--07:26} & 361.83  & 3.71  & 8.84  & 18.29 & 39.01 & 0.72 & 1.02 & 2.51 & N & C \\
13 & \makecell{2021-01-13 \\ 03:50--03:54} & 248.68  & 2.15  & 7.67  & 9.61  & 11.51 & 0.80 & 0.76 & 1.08 & B & C \\
14 & \makecell{2021-01-13 \\ 03:05--03:13} & 455.92  & 5.74  & 6.35  & 14.00 & 20.49 & 1.46 & 1.18 & 2.03 & N & S \\
15 & \makecell{2021-01-13 \\ 05:16--05:22} & 373.03  & 6.58  & 8.23  & 19.24 & 45.23 & 2.81 & 2.28 & 3.71 & N & S \\
16 & \makecell{2021-01-13 \\ 05:59--06:19} & 1202.49 & 10.05 & 44.20 & 14.95 & 44.20 & 3.98 & 2.76 & 5.50 & N & C \\
17 & \makecell{2021-04-07 \\ 06:24--06:32} & 457.88  & 6.82  & 20.54 & 15.30 & 57.90 & 0.46 & 0.59 & 1.67 & N & S \\
18 & \makecell{2021-04-07 \\ 06:35--06:43} & 457.77  & 4.90  & 23.92 & 12.18 & 48.93 & 1.32 & 1.00 & 2.03 & N & C \\
19 & \makecell{2021-04-07 \\ 08:24--08:32} & 622.03  & 6.34  & 7.18  & 13.66 & 27.42 & 0.84 & 0.75 & 1.32 & B & C \\
20 & \makecell{2021-04-07 \\ 06:12--06:19} & 416.32  & 5.02  & 12.76 & 14.55 & 22.88 & 0.69 & 0.76 & 1.91 & N & C \\
21 & \makecell{2021-04-07 \\ 08:31--08:39} & 456.01  & 2.51  & 11.72 & 9.34  & 19.49 & 0.64 & 0.63 & 1.19 & B & C \\
22 & \makecell{2021-04-14 \\ 05:41--05:50} & 540.24  & 4.90  & 7.92  & 10.28 & 21.32 & 0.96 & 1.00 & 2.03 & B & C \\
23 & \makecell{2021-04-14 \\ 05:44--05:59} & 914.01  & 9.33  & 5.96  & 11.70 & 28.70 & 2.29 & 1.16 & 3.23 & B & S \\
24 & \makecell{2021-04-14 \\ 06:28--06:39} & 664.78  & 10.29 & 9.92  & 16.10 & 24.39 & 3.26 & 2.50 & 4.90 & B & S \\
25 & \makecell{2021-04-14 \\ 08:02--08:12} & 581.79  & 7.18  & 6.01  & 14.34 & 27.06 & 2.42 & 2.85 & 4.55 & B & S \\
26 & \makecell{2021-04-14 \\ 06:55--07:05} & 623.23  & 7.42  & 12.35 & 13.01 & 28.17 & 1.11 & 0.87 & 1.44 & N & S \\
27 & \makecell{2021-04-14 \\ 07:08--07:16} & 457.12  & 6.10  & 17.52 & 16.11 & 29.37 & 0.96 & 0.89 & 1.56 & N & S \\
28 & \makecell{2022-11-08 \\ 04:26--04:40} & 793.51  & 6.34  & 10.50 & 8.36  & 14.56 & 1.20 & 0.61 & 1.91 & B & S \\
29 & \makecell{2022-11-08 \\ 04:28--04:32} & 288.59  & 3.83  & 16.42 & 14.36 & 16.42 & 0.90 & 0.71 & 1.32 & B & C \\
30 & \makecell{2022-11-08 \\ 06:49--06:59} & 577.39  & 4.43  & 5.19  & 9.48  & 14.68 & 1.41 & 1.02 & 1.91 & B & S \\
31 & \makecell{2022-11-08 \\ 04:57--05:04} & 432.77  & 3.83  & 15.29 & 10.06 & 16.20 & 0.57 & 0.56 & 0.96 & B & C \\
32 & \makecell{2023-09-06 \\ 01:34--01:43} & 496.04  & 4.90  & 8.72  & 10.80 & 18.10 & 0.70 & 0.93 & 1.79 & N & S \\
33 & \makecell{2023-09-06 \\ 01:39--01:52} & 744.06  & 10.77 & 8.97  & 15.32 & 29.16 & 0.91 & 0.88 & 1.67 & N & S \\
34 & \makecell{2023-09-06 \\ 01:54--02:07} & 744.06  & 9.81  & 11.42 & 17.14 & 26.72 & 2.14 & 1.69 & 3.71 & B & S \\
\enddata

\tablecomments{
$\mathrm{Lifetime}$ denotes the total duration of the plume event, defined as the time interval between the first and last frames.
$\mathrm{Height}$ represents the vertical displacement of the plume, calculated as the difference between the maximum plume heights measured in the first and last frames.
$\mathrm{Velocity}_{\rm ini}$, $\mathrm{Velocity}_{\rm mean}$, and $\mathrm{Velocity}_{\rm max}$ denote the initial, mean, and maximum plume velocities, respectively.
$\mathrm{Width}_{\rm ini}$ denotes the mean plume width in the first frame, whereas $\mathrm{Width}_{\rm mean}$ and $\mathrm{Width}_{\rm max}$ represent the event mean width and event maximum width over the entire plume lifetime, respectively.
$\mathrm{Loc.}$ indicates the formation location of the plume, where ``B'' and ``N'' represent bubble and non-bubble plumes, respectively.
$\mathrm{Curv.}$ indicates the trajectory morphology, where ``C'' and ``S'' denote curved and quasi-straight plume trajectories, respectively, based on the median mean-curvature threshold (0.4~$\mathrm{Mm}^{-1}$).
}

\end{deluxetable*}

\section{Result and Discussion}
\subsection{Overall Distribution of Physical Properties of Plumes and Their Evolution Properties}

\begin{figure*}[h]
\centering
\includegraphics[width=1.\textwidth]{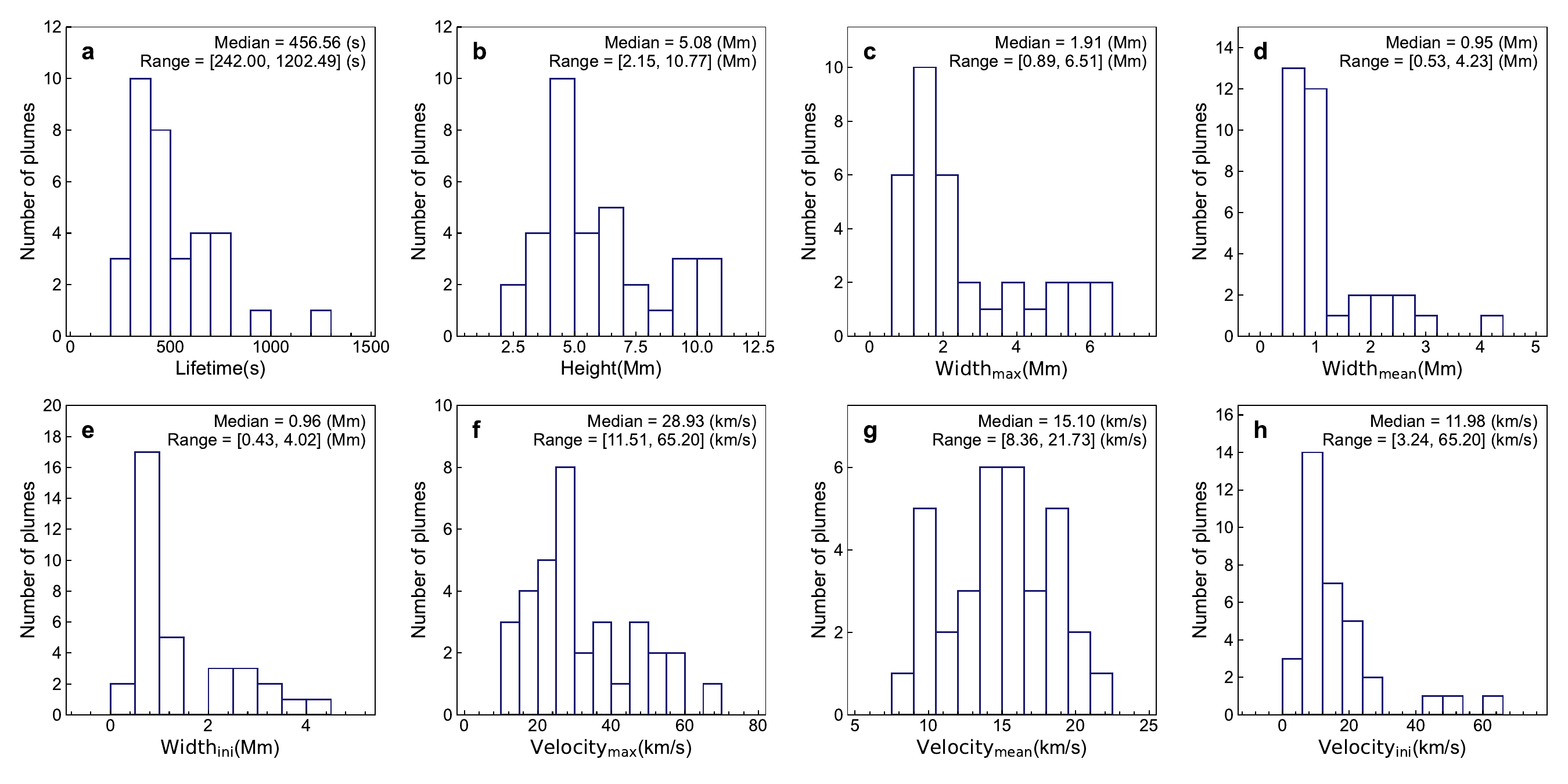}
\caption{Statistical distributions of plume dynamical and geometrical parameters, including lifetime(a), vertical displacement (b), width (c--e), and velocity (f--h). The histograms represent the distributions of the overall sample. The median values and parameter ranges are marked in each panel.}
\label{fig:Figure2}
\end{figure*}

As shown in Figure~\ref{fig:Figure2}, both the dynamical and geometrical parameters exhibit markedly asymmetric distributions. The plume lifetime is primarily concentrated in the range of 300--700~s, with a median value of approximately 450~s. Most events belong to the short-lived category, although a few long-lived plumes are also observed (Figure~\ref{fig:Figure2}(a)). In terms of geometrical parameters, the plume vertical displacement (Figure~\ref{fig:Figure2}(b)) range from 2.15 to 10.77~Mm, with most events concentrated between 3 and 7~Mm. The maximum width ($\text{Width}_{\max}$), mean width ($\text{Width}_{\text{mean}}$), and initial width ($\text{Width}_{\text{ini}}$) range from 0.89 to 6.51~Mm, 0.53 to 4.23~Mm, and 0.43 to 4.02~Mm, respectively, with corresponding median values of 1.91~Mm, 0.95~Mm, and 0.96~Mm. Their distributions are mainly concentrated in the ranges of 1--3~Mm, 0.5--1.5~Mm, and 0.5--1.5~Mm, respectively. Nevertheless, a small number of plumes occupy the high-width tail of the distribution. The mean velocity is mainly distributed between 10 and 20~$\mathrm{km~s^{-1}}$, with a median value of about 15~$\mathrm{km~s^{-1}}$ and relatively low dispersion. In contrast, the distributions of both the maximum and initial velocities exhibit high-value tails, extending to approximately 65~$\mathrm{km~s^{-1}}$, with their values mainly concentrated in the ranges of 15--40~$\mathrm{km~s^{-1}}$ and 5--25~$\mathrm{km~s^{-1}}$, respectively. 

The characteristic ranges of several key physical parameters derived from our sample of 34 plume events, including the mean velocity (8.36--21.73~$\mathrm{km~s^{-1}}$), maximum width (0.89--6.51~$\mathrm{Mm}$), and lifetime (242.00--1202.49~s), are generally consistent with those reported by \citet{2010ApJ...716.1288B} for five representative plumes (13--17~$\mathrm{km~s^{-1}}$, 2--6~$\mathrm{Mm}$, and 400--890~s, respectively). The agreement between the two studies supports the reliability of the statistical properties derived in this work. Meanwhile, the larger sample size enables a more comprehensive characterization of the full parameter space, revealing broader distributions and more diverse evolutionary characteristics than could be inferred from a limited number of case studies. These broad, skewed parameter distributions and the associated diverse evolutionary behaviors, particularly the pronounced asymmetry and the extended high-value tails observed in the velocity and width distributions, indicate that there is event-to-event variability in the geometrical morphology and dynamical evolution of plumes. Therefore, the formation and subsequent evolution of plumes cannot be explained by a single, idealized physical mechanism. Such diversity may arise from different triggering conditions for plume formation and may also be influenced by variations in the local magnetic configuration, the surrounding plasma environment, and other external perturbations.

\begin{figure*}[h]
\centering
\includegraphics[width=0.95\textwidth]{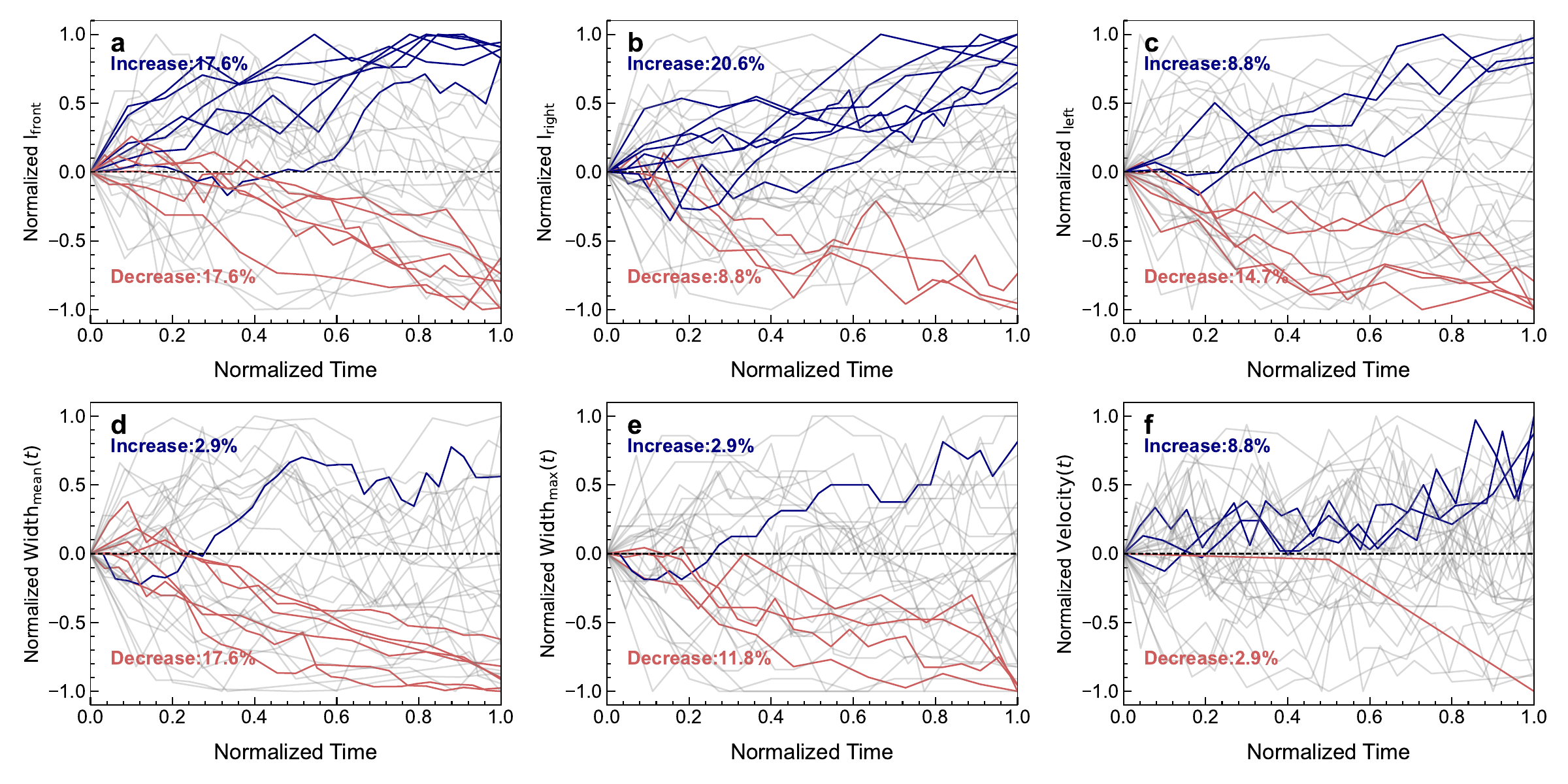}
\caption{Normalized temporal evolution of plume physical parameters. Evolutionary characteristics are shown for the radiation intensity time series in the plume-front (a), right flank (b), and left flank (c) regions, as well as the frame mean width time series, frame maximum width time series, and velocity time series (d--f). All parameters are normalized to their respective initial values and aligned onto a common normalized time axis. Each curve represents an individual plume event and is classified according to its overall evolutionary trend into three categories: increasing (blue), decreasing (red), and no significant trend (gray). The percentages of events exhibiting distinct evolutionary trends are indicated in each panel.}
\label{fig:Figure3}
\end{figure*}

In Figure~\ref{fig:Figure3}, we compared the evolutionary trends of physical parameters among different plume events. The results show that none of the parameters exhibit a dominant monotonic trend, as the majority of events are characterized by non-monotonic fluctuations (indicated by the gray curves). Furthermore, among the few plumes exhibiting monotonic trends, there is a clear directional asymmetry, with a strong preference for either an increasing or decreasing trend. For the radiation intensity in the plume-front region (Figure~\ref{fig:Figure3}(a)), the fractions of events exhibiting increasing and decreasing trends are identical and both low (17.6\%). In the left and right flank regions (Figure~\ref{fig:Figure3}(b) and (c)), the fractions of events with increasing trends in radiation intensity are 8.8\% and 20.6\%, while the corresponding fractions with decreasing trends in radiation are 14.7\% and 8.8\%. This directional asymmetry suggests that the two sides of a plume may be embedded in different background environments, such as asymmetric magnetic fields or plasma conditions, or that the plume motion may contain a lateral velocity component. 

The evolution of the frame mean and frame maximum widths (Figure~\ref{fig:Figure3}(d) and (e)) is predominantly characterized by random fluctuations without a clear monotonic trend. Nevertheless, decreasing trends are more common than increasing trends, with 17.6\% and 11.8\% of events exhibiting contraction in the frame mean and frame maximum width time series, respectively, compared with only 2.9\% showing increasing trends for both width parameters. This behavior may result from compression by the surrounding plasma or confinement by the background magnetic field during the evolution of plumes. The velocity time series (Figure~\ref{fig:Figure3}(f)) is also dominated by fluctuating behavior. Only about 8.8\% of the events display a sustained acceleration trend, while 2.9\% of the events exhibit a persistent deceleration trend. This finding suggests that plume dynamics are not driven solely by the buoyant rise of low-density plasma. Additional upward driving forces may continue to act during the evolution of plumes. Such forces could originate from enhanced magnetic pressure associated with local magnetic field evolution, energy release through magnetic reconnection, or possibly thermal pressure gradients generated by localized heating \citep{2012ApJ...761....9D,2022A&A...659A..76W}. At the same time, the upward motion may also be moderated by opposing forces, such as magnetic tension or interaction with the surrounding prominence plasma. Overall, the evolution of plumes does not follow a single dynamical pattern but instead exhibits diverse behaviors, suggesting that their formation and evolution could be shaped by a combination of external disturbances acting on the prominence, alongside intrinsic magnetic field and plasma dynamics.

\subsection{Relationships between Key Parameters of Plumes}

\begin{figure*}[htbp]
\centering
\includegraphics[width=0.95\textwidth]{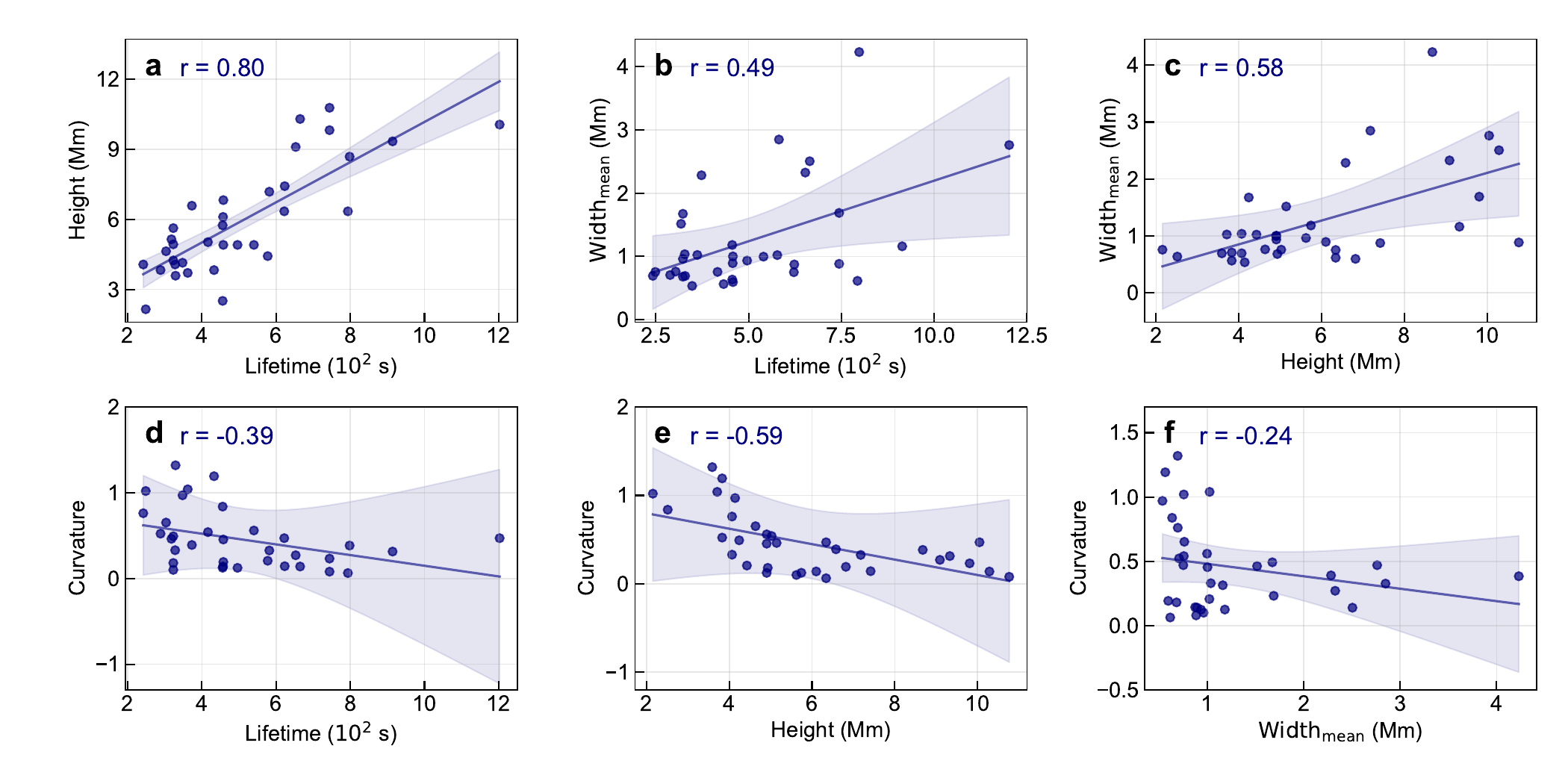}
\caption{Correlation analysis and linear regression fits among the key parameters of plumes.Upper panels: (a) Height vs. Lifetime, (b) $\mathrm{Width}_{\rm mean}$ vs. Lifetime, and (c) $\mathrm{Width}_{\rm mean}$ vs. Height. Lower panels: (d) Curvature vs. Lifetime, (e) Curvature vs. Height, and (f) Curvature vs. $\mathrm{Width}_{\rm mean}$. In each panel, the solid blue line represents the linear regression fit, with the light blue shaded area denoting the $\pm 1\sigma$ standard error band. The $r$ value indicates the Pearson correlation coefficient.}
\label{fig:Figure4}
\end{figure*}

Building upon the above results, Figure~\ref{fig:Figure4} further investigates the relationships among the principal parameters, such as lifetime, vertical displacement, velocity, mean width, and trajectory curvature. After checking the relationships between all parameter pairs, we found that lifetime, vertical displacement, and mean width are all positively correlated with one another. Among these, the strongest correlation is found between lifetime and vertical displacement ($r=0.80$), indicating that plumes with longer lifetimes generally attain greater upward extents during their evolution. Both the $\mathrm{Width}_{\rm mean}$-Lifetime and $\mathrm{Width}_{\rm mean}$-Height relationships exhibit moderate positive correlations, suggesting that the temporal and spatial scales of plume evolution are not independent but are intrinsically coupled. 

In addition, the propagation trajectory, one of the most direct manifestations of the spatiotemporal evolution of plumes, is also analyzed. We find that as plumes rise into the prominence, their trajectories exhibit varying degrees of bending and deflection, indicating substantial differences in their evolutionary behavior after entering the prominence body. Then we further investigate the relationship between trajectory curvature and plume dynamical properties. Figures~\ref{fig:Figure4}(d)--(f) reveal that curvature is negatively correlated with both plume lifetime ($r=-0.39$) and vertical displacement ($r=-0.59$): plumes with lower trajectory curvature tend to have longer lifetimes and achieve greater upward displacements. These findings suggest that trajectory curvature not only characterizes the geometric morphology of plume propagation paths but is also closely related to other dynamical properties of plume evolution.

\begin{figure*}[htbp]
\centering
\includegraphics[width=1\textwidth]{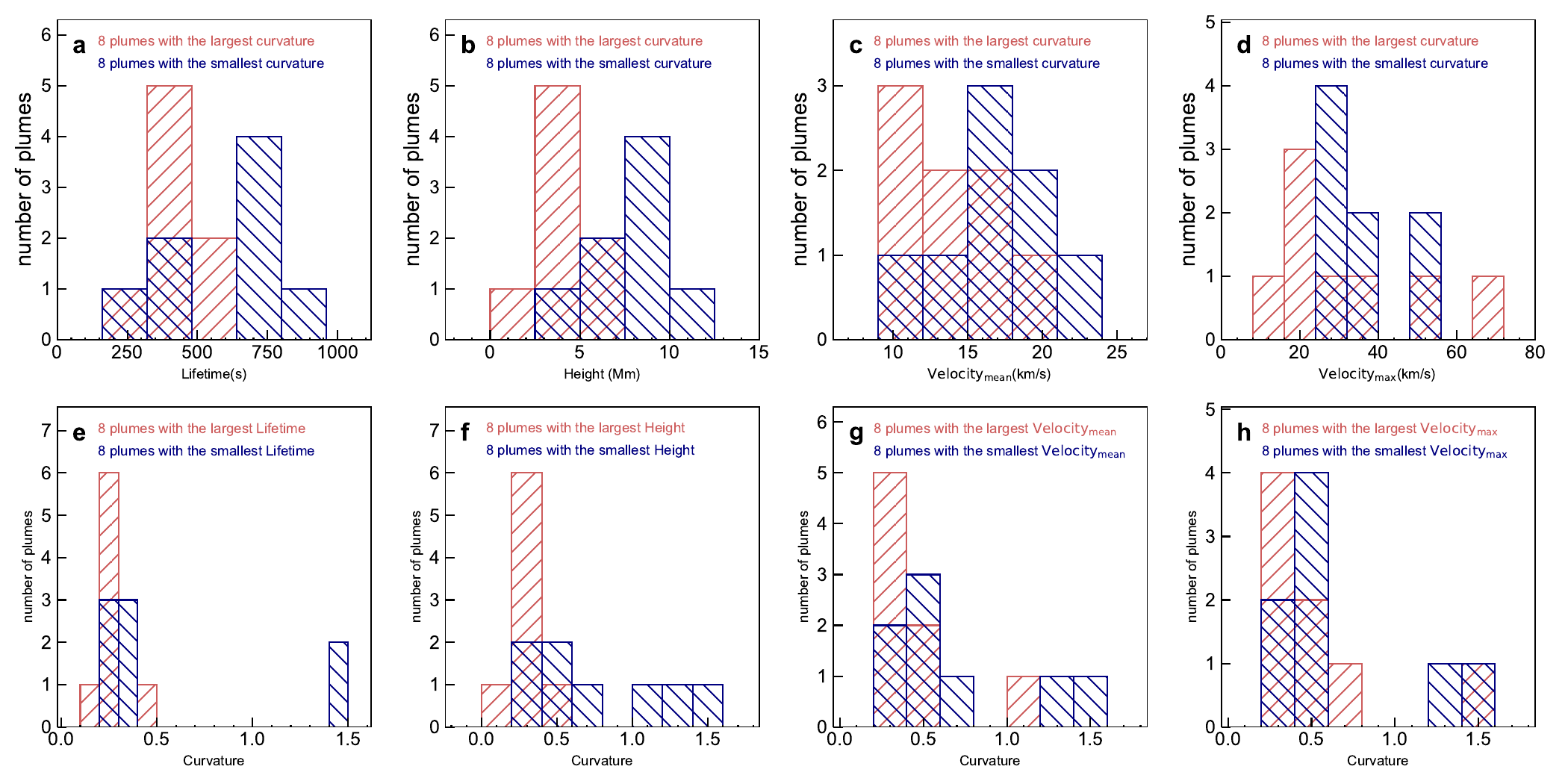}
\caption{Relationship between plume trajectory curvature and plume lifetime (a), vertical displacement (b), mean velocity (c), and maximum velocity (d). The upper panels use trajectory curvature as the selection criterion and compare the distributions of lifetime (a), vertical displacement (b), and velocity (c--d) for the eight plume events with the highest and lowest curvature values. The lower panels reverse the selection criterion by selecting the eight plume events with the highest and lowest values of lifetime (e), vertical displacement (f), and velocity (g--h), respectively, and comparing their curvature distributions.}
\label{fig:Figure5}
\end{figure*}

Based on the above results, we further selected two groups of plume events with extreme trajectory curvatures—one consisting of plumes with the largest curvatures and the other consisting of plumes with the smallest curvatures—for comparative analysis. Figures~\ref{fig:Figure5}(a)--(d) present the distributions of plume lifetime, vertical displacement, mean velocity, and maximum velocity for these two groups. It can be seen that low-curvature plumes generally exhibit larger values of lifetime, vertical displacement, mean velocity, and maximum velocity. Subsequently, we found that Plumes with larger lifetimes, vertical displacements, and velocities generally exhibit lower trajectory curvatures (Figures~\ref{fig:Figure5}(e)--(h)). These findings suggest that low-curvature plumes experience less resistance from the surrounding prominence environment during their ascent, enabling them to maintain relatively higher velocities and persist for longer durations. High-curvature plumes, on the other hand, appear to be more strongly influenced by the surrounding prominence structure, resulting in reduced velocities and shorter evolutionary timescales.

\begin{figure*}[h]
\centering
\includegraphics[width=0.8\textwidth]{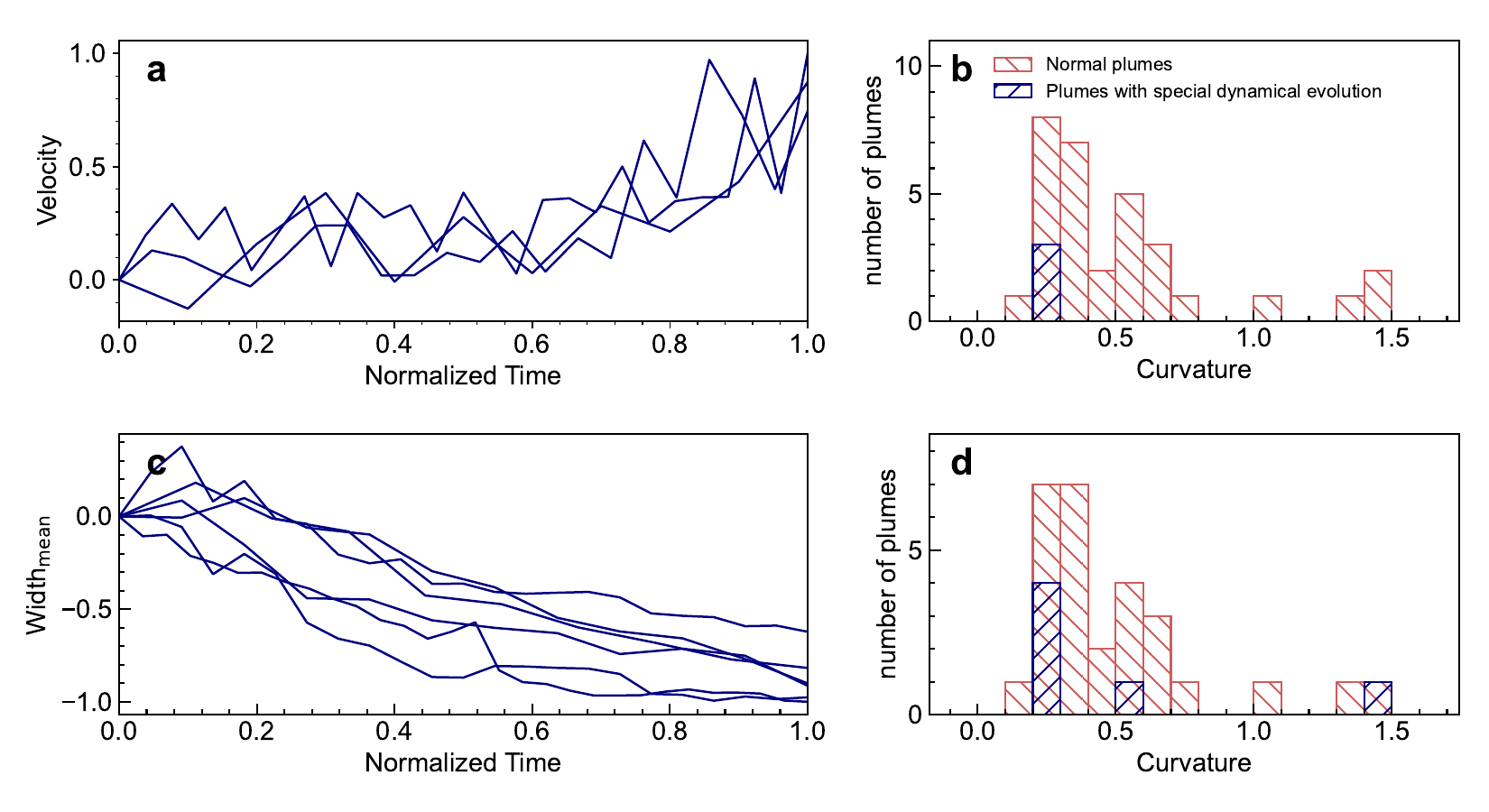}
\caption{Normalized temporal evolution and curvature distributions of two special plume categories. Panels (a) and (c) show the normalized temporal evolution of plumes exhibiting continuously increasing velocity and gradually decreasing frame mean width, respectively. Panels (b) and (d) compare the curvature distributions of these two categories (blue hatched bars) with those of the remaining plume sample (red hatched bars).}
\label{fig:Figure6}
\end{figure*}

To further investigate whether plume trajectory curvature is also related to the evolutionary behaviors of other parameters, we focused on plume events exhibiting distinct evolutionary patterns identified in Figure~\ref{fig:Figure3}. In particular, we examined whether these events show a significant preference for specific curvature ranges (see Figure~\ref{fig:Figure6}). Two categories of plumes with distinctive evolutionary characteristics were selected for analysis: one exhibiting a continuous increase in velocity during evolution and the other showing a gradual decrease in mean width over time. The results show that plumes with gradually increasing velocity and gradually decreasing mean width are mainly concentrated in the low curvature region.

The pronounced low-curvature characteristic of plumes with sustained acceleration is consistent with physical expectations, as low-curvature plumes are likely to experience less resistance from the surrounding prominence environment during their ascent. Additionally, we calculated that for plumes exhibiting continuous mean width contraction, their initial widths are clustered in a relatively large range, between 2.3 and 3.3 Mm. We speculate that they may possess comparatively stable intrinsic magnetic structures, such as plumes formed through the eruption of mini-filaments \citep{2021ApJ...923L..10C,2022A&A...659A..76W,2024ApJ...970..110G}. Because such plumes are associated with magnetic structures originating from eruptions in the underlying atmosphere, their subsequent evolution may be less affected by the relatively weak magnetic fields within the prominence environment. As a result, they can evolve in a more stable manner and are therefore more likely to exhibit trajectories with low curvature. These results suggest that trajectory curvature reflects the interplay between external environmental influences and the intrinsic magnetic stability of the plume structure. 

As these plumes ascend, their motion inevitably perturbs or even interacts with the magnetic field and plasma environment of the surrounding prominence. These perturbations could excite small-scale transverse or longitudinal oscillations in adjacent prominence threads \citep{2012A&A...542A..52Z,2014ApJ...795..130S,
2022SCPMA..6539611L,2024ApJ...965L..28W,2025ApJ...981..139Y}, which warrant further investigation. Future studies combining high-resolution observations, magnetic field measurements, and 3D MHD simulations are desirable to further elucidate the coupling between plume dynamics and the overall dynamical stability of solar prominences.

It is worth noting that these results rely on single-viewpoint observations, which only reveal the morphology and dynamics of plumes projected to the two-dimensional plane of sky (POS). As a result, the key parameters such as the width and trajectory curvature of plumes will inherently suffer from Line-of-Sight (LOS) projection effects. However, after performing a LOS projection effect analysis, we demonstrated that the LOS projection effect on the trajectory curvature is high only for a small subset of specific LOS configurations. For the majority of projection angles, the variation in curvature remains relatively modest and does not drastically alter the measured values. Consequently, in our ensemble statistics, omitting this projection effect may not exert a significant impact on the overall statistical trends. We would also like to emphasize one point: although multi-viewpoint, high spatial and temporal resolution H$\alpha$ observations are currently unavailable—preventing an in-depth analysis of the three-dimensional dynamical evolution of the prominence plume—in the future, we can combine high-resolution, multi-viewpoint ultraviolet data from the Solar Orbiter spacecraft to carry out such investigations.

\subsection{Initial Conditions and Possible Driving Mechanisms of Plumes}

The aforementioned results indicate that different plumes exhibit vastly diverse dynamical evolution characteristics. To analyze the origin of this diversity, it is necessary to further explore the initial states of the plumes as well as their potential driving mechanisms. Regarding the initial state of the plumes, our statistical results show that approximately 60\% of plumes originate at bubble boundaries, whereas the remaining 40\% form in regions not associated with bubble boundaries. This indicates that plume formation models based solely on bubble boundaries cannot explain all plume events. Therefore, the occurrence location of a plume (i.e., whether it is associated with a bubble structure) constitutes the primary feature for evaluating differences in its initial state. In addition, the initial width and initial velocity also harbor crucial information regarding the initial state of the plumes.

As for the driving mechanisms of the plumes, previous studies suggest two major categories: one comprises MHD instability mechanisms, represented by the RT instability and Kelvin–Helmholtz–Rayleigh–Taylor (KH-RT) instability \citep{2010ApJ...716.1288B,2010SoPh..267...75R,2012ApJ...746..120H,2012ApJ...756..110H,2018RvMPP...2....1H}; the other emphasizes the evolutionary role of the prominence magnetic field induced by perturbations originating below the prominence, such as the magnetic deformation or pressure enhancement driven by an unstable rising mini-filament or magnetic flux rope beneath the prominence, even magnetic reconnection at the bubble boundaries \citep{2012ApJ...761....9D,2014A&A...567A.123G,2015ApJ...814L..17S,2021ApJ...923L..10C,2022A&A...659A..76W,2024ApJ...970..110G}. Past studies have pointed out that localized brightening at bubble boundaries could be key observational evidence supporting the latter category of mechanisms. \cite{2021ApJ...911L...9G} found that during the collapse of Type-II bubbles and the subsequent formation of plumes, conspicuous localized brightening appeared at the bubble boundary corresponding to the plume generation site, which might be attributed to weak magnetic reconnection or plasma piling. \cite{2021ApJ...923L..10C} also observed brightening at bubble edges and suggested that it could be related to energy release (such as magnetic reconnection) associated with magnetic flux ropes or, alternatively, enhanced coronal emission revealed after the flux rope cleared out the cold plasma. These studies suggest that boundary brightening may be associated with flux rope interactions or magnetic reconnection processes at bubble boundaries. Therefore, this study adopts ``whether a conspicuous precursor brightening exists at the trigger location" as the key criterion to distinguish between different driving mechanisms, thereby performing a comparative analysis of plumes with different initial states.

To objectively determine precursor brightening events, this study selects the frame immediately preceding plume appearance as the reference time ($t_{\rm ref}$) and defines a small area at the plume front as the trigger location. Within the 3.5 minutes before plume appearance, the maximum radiative flux is taken as the peak brightening intensity ($I_{\rm peak}$). The peak lead time is defined as $\Delta t=t_{\rm ref}-t_{\rm peak}$, where $t_{\rm peak}$ is the occurrence time of $I_{\rm peak}$. Within 3 minutes before the peak, the minimum flux is identified as the valley, and the 6 minutes preceding the valley are used as the background interval to calculate the mean ($\mu$) and standard deviation ($\sigma$). A significant precursor brightening is identified when $I_{\rm peak}>\mu+2\sigma$ and $\Delta t\le2$ min. After excluding Cases 12, 14, and 32 due to insufficient background coverage, 31 valid plume events remained for analysis.

\begin{figure*}[h]
\centering
\includegraphics[width=1.\textwidth]{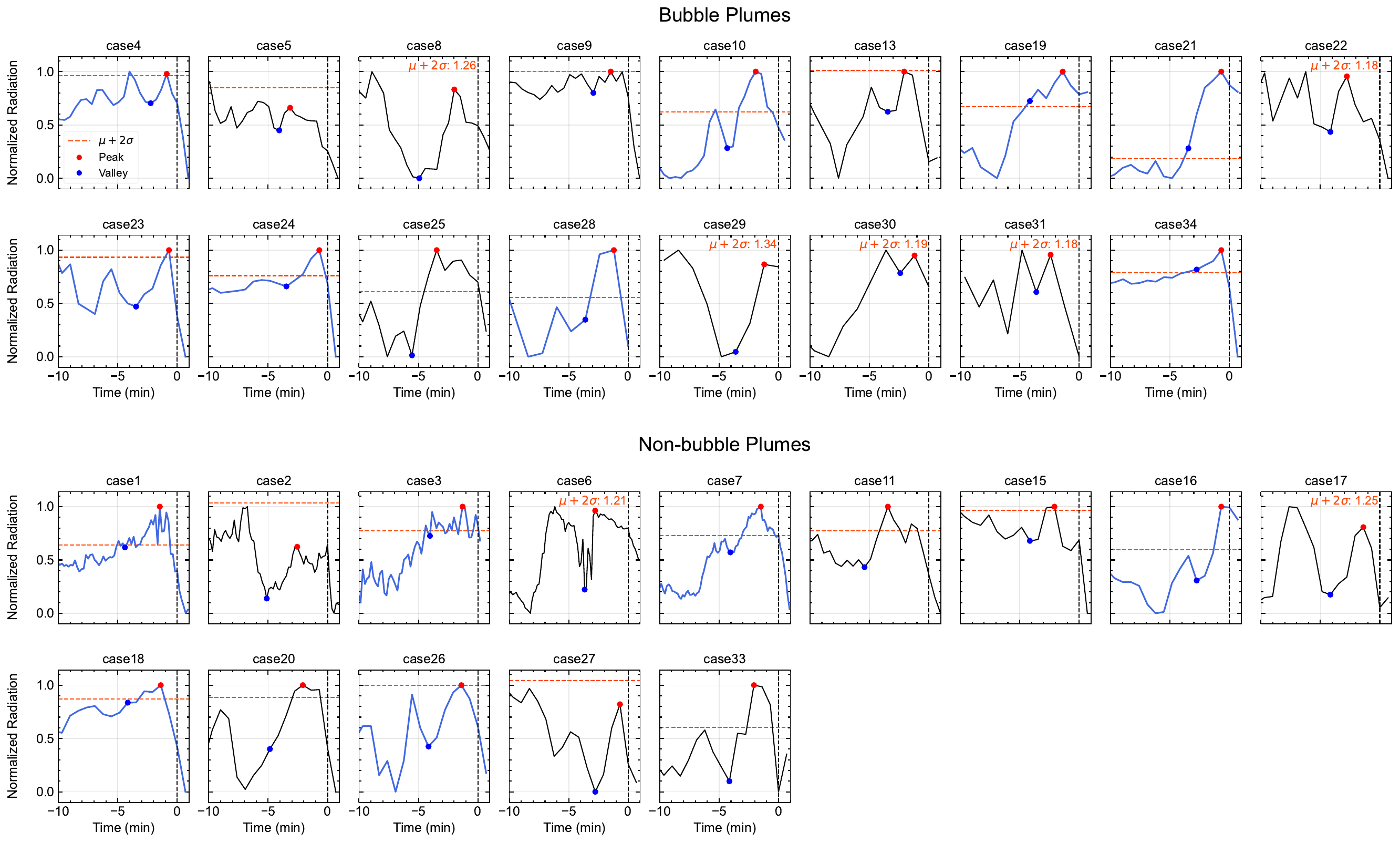}
\caption{Normalized radiation intensity curves of the trigger location for bubble and non-bubble plumes. The upper panels show bubble plumes, while the lower panels show non-bubble plumes. The blue curves denote plume events exhibiting brightening behavior. Red and blue points indicate the locations of brightness peaks and local minima, respectively. The horizontal red dashed lines indicate the corresponding $\mu+2\sigma$ levels. The dashed vertical line marks the plume onset time (t = 0).}
\label{fig:Figure7}
\end{figure*}

\begin{figure*}[h]
\centering
\includegraphics[width=1.\textwidth]{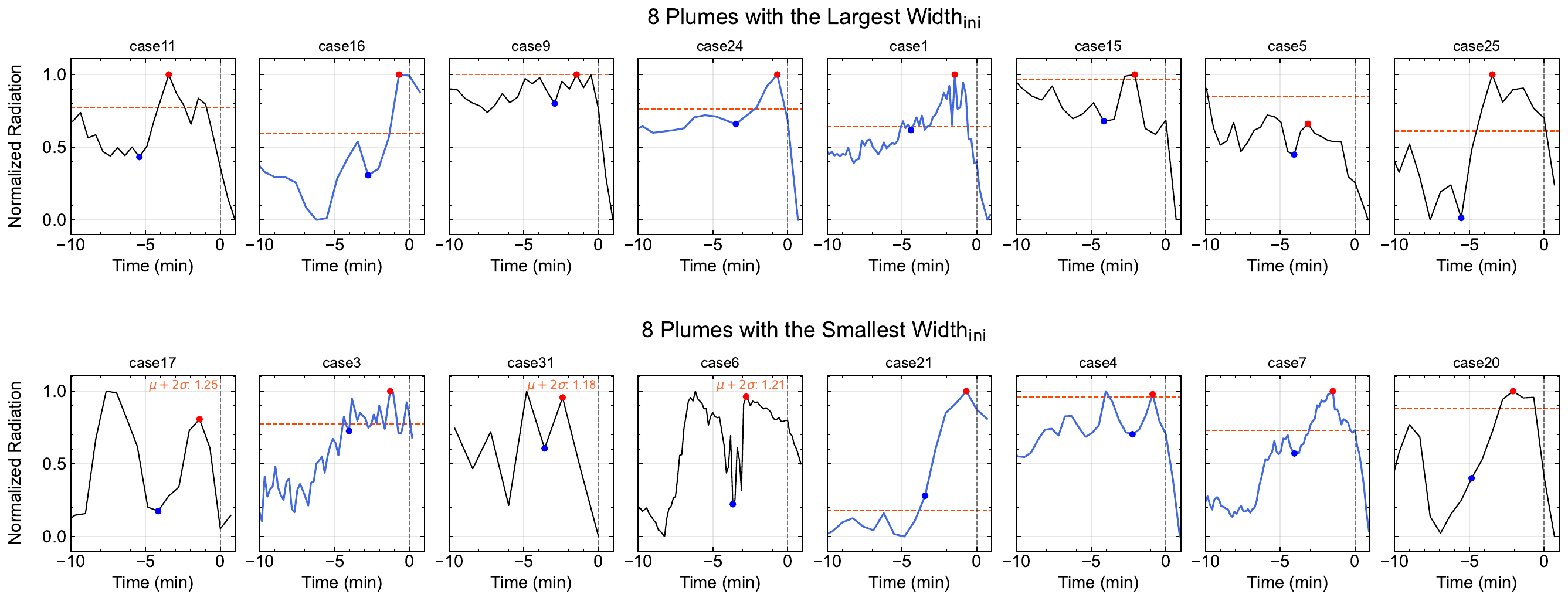}
\caption{Similar to Figure~\ref{fig:Figure7}, but for bubbles with different initial widths. The upper panels show the eight plume events with the largest initial widths, while the lower panels show the eight plume events with the smallest initial widths.}
\label{fig:Figure8}
\end{figure*}

\begin{figure*}[h]
\centering
\includegraphics[width=1.\textwidth]{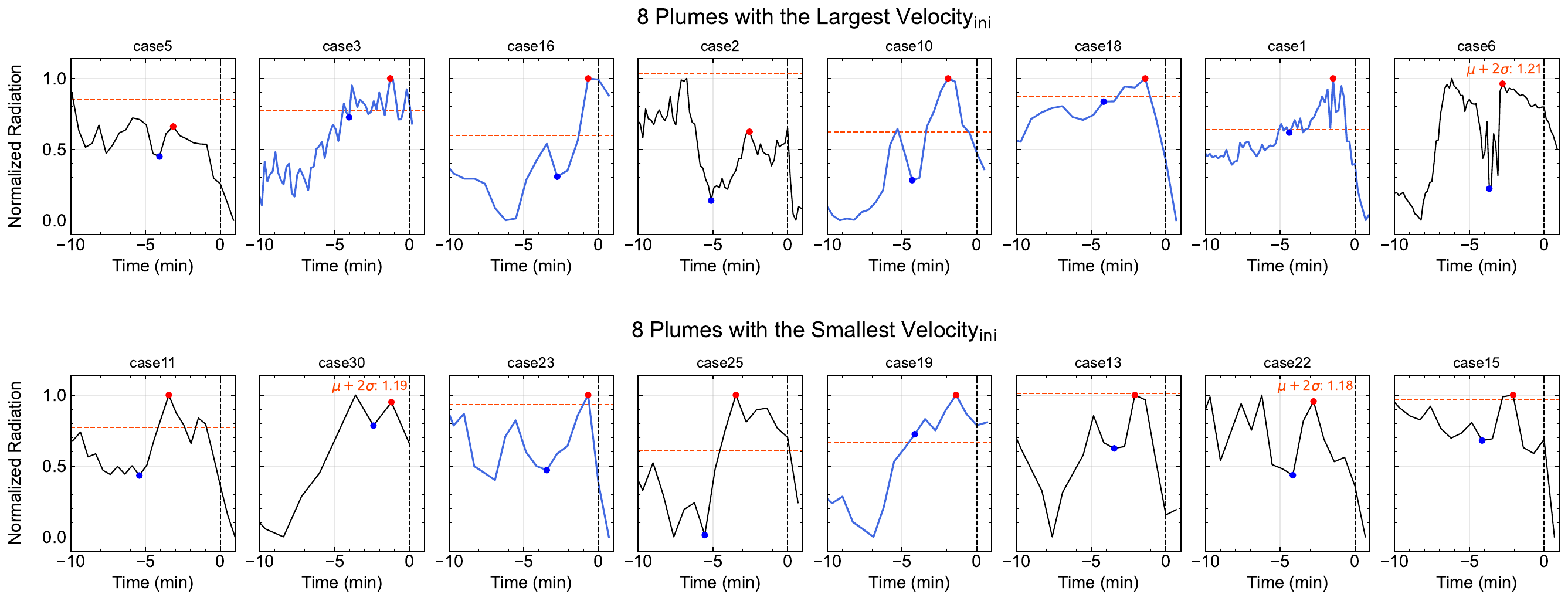}
\caption{Similar to Figure~\ref{fig:Figure7}, but for bubbles with different initial velocities. The upper panels show the eight plume events with the highest initial velocities, while the lower panels show the eight plume events with the lowest initial velocities.}
\label{fig:Figure9}
\end{figure*}

First, we classified the 31 plumes into two categories based on whether their trigger locations were situated at bubble boundaries (Figure~\ref{fig:Figure7}). The results reveal that the proportion of samples accompanied by precursor brightening shows virtually no difference between the two groups (47.1\% for bubble-plume events versus 42.9\% for non-bubble-plume events). This indicates that the appearance of localized brightening prior to the plume onset is not restricted by its initial location; that is, the brightening-related physical mechanisms (such as localized magnetic reconnection, instabilities, or flux rope lifting) can occur at any location beneath the prominence to subsequently generate plumes. This is consistent with the physical picture proposed by \citep{2024ApJ...970..110G}: Type-II bubbles triggered by mini-filament eruptions can manifest as plume structures under specific LOS viewing angles and projection effects, regardless of whether a bubble is visibly present. The authors also found that during the collapse and plume-formation process of such Type-II bubbles, their leading edges exhibit significant localized brightening observationally due to energy release or severe compression. As a result, we speculate that some of the ``non-bubble region" plumes observed in this study may inherently be such transient bubbles directly driven by mini-filament eruptions; however, because the axis of the erupting mini-filament is parallel to the LOS, they manifest as rapidly rising plumes under the projection effect.

Subsequently, we ranked the physical parameters of the samples and extracted two sets of extreme samples with the largest and smallest values (8 samples per group) for comparison, focusing on the initial width (Figure~\ref{fig:Figure8}) and initial velocity (Figure~\ref{fig:Figure9}). The results show no significant difference in the proportion of samples accompanied by precursor brightening between the extreme groups of different initial widths. In contrast, among the 8 plumes in the high-initial-velocity group, 5 samples (62.5\%) are accompanied by distinct precursor brightening, whereas in the low-velocity group of 8 plumes, only 2 samples (25\%) exhibit brightening characteristics. We conjecture that plumes with high initial velocities are more likely to be dominated by the second category of driving mechanisms (i.e., magnetic reconnection or mini-filament eruption), where the plumes are endowed with higher initial kinetic energy. The significant correlation between precursor brightening and initial velocity, but not initial width, may arise because plume width is susceptible to two-dimensional projection effects along the LOS. As \citet{2024ApJ...970..110G} noted, Type-II bubbles from mini-filament eruptions can appear as plumes under certain viewing angles. Initial velocity, by contrast, is less affected by projection, thus more faithfully reflecting the energy release at the trigger location.

Although prominence bubbles and erupting mini-filaments extend over tens of Mm \citep{2010ApJ...716.1288B,2015ApJ...814L..17S,2021ApJ...911L...9G,2024ApJ...970..110G}, the dark upward plumes have markedly smaller spatial scales, with widths of approximately 0.5--4.0 Mm reported in the 3D MHD numerical simulations of \cite{2012ApJ...746..120H}, which is also consistent with our observed initial plume widths (0.4--4.0 Mm). This indicates that the magnetic reconnection driving plumes may occur on substantially smaller spatial domains ($\sim 1\text{ Mm}$), corresponding to the local regions of direct interaction between an erupting mini-filament and the overlying prominence or bubble. Guided by the lower limit of observed plume scales, we selected local regions with dimensions of approximately 0.3--0.4 Mm for light-curve extraction of precursor brightenings during plume formation.

\section{Summary}
Based on high-resolution NVST H$\alpha$ observation data obtained from 2013 to 2025, this study constructed a catalog of prominence samples containing 201 prominences and a catalog of plume events comprising 137 plumes, from which 34 typical plumes with complete evolutionary processes were investigated in detail. Utilizing uniform image processing and parameter extraction methods, the morphological and kinematic parameters of these plumes were obtained. A systematic analysis of these plumes was then conducted across multiple dimensions, including statistical distributions, temporal evolution characteristics of physical parameters, parameter correlations, categorical comparative characteristics, formation environments, and potential triggering mechanisms. The main results are as follows:

\begin{enumerate}
    \item Plume lifetimes are primarily concentrated in the range of 300--700~s, with vertical displacements mostly between 3.0 and 7.0~Mm. The mean widths and velocities are concentrated in the ranges of 0.5--1.5~Mm and 10--20~km~s$^{-1}$, respectively. Several parameters exhibit pronounced high-value tails. The temporal evolution of plume physical parameters is generally characterized by the absence of clear monotonic trends. The wide parameter distribution and diverse evolutionary patterns imply that plumes are potentially driven by multiple mechanisms, shaped by a combination of external disturbances, intrinsic magnetic fields, and plasma dynamics throughout their evolution.
    
    \item The lifetime, vertical displacement, and mean width exhibit significant pairwise positive correlations. This suggests that the temporal and spatial scales of plumes do not evolve independently but rather possess an intrinsic coupling mechanism.
    
    \item Plume trajectory curvature correlates negatively with lifetime, vertical displacement, and velocity. Plumes with continuously increasing velocity and decreasing width concentrate in low-curvature regions. This makes curvature a key geometric indicator: low-curvature trajectories may signify more stable internal magnetic structures or less resistance and perturbation from the background prominence.
    
    \item The identical proportion of precursor brightening between bubble and non-bubble groups indicates that plume initiation is independent of macro-geometric positioning or bubble structures. Instead, diverse localized mechanisms can trigger plumes. Some non-bubble plumes may inherently be transient bubbles driven by eruptive mini-filaments or magnetic flux ropes, manifesting as plumes due to projection effects.
    
    \item Extreme group analysis reveals a higher proportion of precursor brightening in high-initial-velocity plumes, suggesting that larger initial velocities may be linked to localized energy release. This supports the idea that certain high-initial-velocity plumes may be driven by magnetic reconnection or flux rope instability, although the involvement of MHD instabilities cannot be ruled out.
\end{enumerate}

In summary, these statistical results demonstrate that prominence plumes are unlikely to be driven by a single, uniform physical mechanism. Instead, their formation and evolution are more likely shaped by a combination of external disturbances acting on the prominence, along with its intrinsic magnetic field and plasma dynamics. Additionally, although MHD instability and magnetic reconnection mechanisms are treated here independently for analytical convenience, they likely operate in tandem and couple mutually in realistic prominence environments during the formation of plumes. Therefore, rather than singling out a dominant mechanism, this study statistically explored environmental differences among plumes, thereby providing new constraints on the diversity of plume formation mechanisms.

It should also be noted that, limited by current observational instrumentation and technical means, this study has the following limitations: First, the stringent selection criteria of plumes adopted in this study restrict the effective sample size and may introduce potential survival biases. Since we focused on plume events with clearly identifiable boundaries and complete evolutionary processes, plumes with smaller scales or lower kinetic energies are less likely included in the present sample. Second, fine-scale structures during the initial triggering phase of plumes and the kinematic details of rapid evolution remain difficult to resolve at current telescope resolutions. Additionally, automated algorithms inevitably introduce errors in boundary identification and parameter extraction due to data-quality constraints. Therefore, higher spatiotemporal resolution observations and numerical simulations are still needed to further clarify the physical nature and formation mechanisms of prominence plumes.

\section*{Acknowledgments}
The authors appreciate the anonymous referee for the constructive comments and valuable suggestions. The data used here are courtesy of the NVST science team. The authors are supported by the National Key R\&D Program of China (2022YFF0503800), the National Natural Science Foundation of China (12273060, 12325303, 12473059, 12373063), the Strategic Priority Research Program of CAS (XDB0560000), the Youth Innovation Promotion Association CAS (2023063), Yunnan Key Laboratory of Solar Physics and Space Science (202205AG070009), Yunnan Fundamental Research Projects (202501AW070002), the Youth Research Special Project of NCUT (2025NCUTYRSP038), Research Startup Fund of NCUT (11005136025XN076-072), China's Space Origins Exploration Program (GJ11020405), Sichuan Science and Technology Program (2025ZNSFC0877), and the Specialized Research Fund for State Key Laboratory of Solar Activity and Space Weather.

\bibliography{sample701}{}

@ARTICLE{2026ApJ..1004...66Z,
       author = {{Zhang}, Junyi and {Hou}, Yijun and {Liu}, Xiaofeng and {Li}, Ting and {Rao}, Shihao and {Qiu}, Ye and {Jin}, HuiPing and {Cai}, Yingjie and {Chen}, Yangrui and {Li}, Chuan},
        title = "{Can We Distinguish the Source Region Location of Filament/Prominence Eruptions from the Sun-as-a-star H{\ensuremath{\alpha}} Spectrum?}",
      journal = {\apj},
         year = 2026,
        month = jun,
       volume = {1004},
       number = {1},
          eid = {66},
        pages = {66},
          doi = {10.3847/1538-4357/ae6c22},
archivePrefix = {arXiv},
       eprint = {2605.10891},
 primaryClass = {astro-ph.SR},
       adsurl = {https://ui.adsabs.harvard.edu/abs/2026ApJ..1004...66Z}
}

@ARTICLE{2025ApJ...993..126L,
       author = {{Liu}, Xiaofeng and {Hou}, Yijun and {Li}, Ying and {Qiu}, Ye and {Li}, Ting and {Cai}, Yingjie and {Rao}, Shihao and {Zhang}, Junyi and {Li}, Chuan},
        title = "{Sun-as-a-star Analysis of the Solar Eruption Source Region Using H{\ensuremath{\alpha}} Spectroscopic Observations of CHASE}",
      journal = {\apj},
         year = 2025,
        month = nov,
       volume = {993},
       number = {1},
          eid = {126},
        pages = {126},
          doi = {10.3847/1538-4357/ae0743},
archivePrefix = {arXiv},
       eprint = {2508.17762},
 primaryClass = {astro-ph.SR},
       adsurl = {https://ui.adsabs.harvard.edu/abs/2025ApJ...993..126L}
}

@ARTICLE{2025ApJ...987...38Z,
       author = {{Zhou}, Chengrui and {Shen}, Yuandeng and {Xia}, Chun and {Liu}, Dongxu and {Tang}, Zehao and {Yao}, Surui},
        title = "{A New Formation Mechanism of Counterstreaming Mass Flows in Filaments and the Doppler Bullseye Pattern in Prominences}",
      journal = {\apj},
         year = 2025,
        month = jul,
       volume = {987},
       number = {1},
          eid = {38},
        pages = {38},
          doi = {10.3847/1538-4357/add32f},
archivePrefix = {arXiv},
       eprint = {2504.14984},
 primaryClass = {astro-ph.SR},
       adsurl = {https://ui.adsabs.harvard.edu/abs/2025ApJ...987...38Z}
}

@ARTICLE{2008ApJ...676L..89B,
       author = {{Berger}, Thomas E. and {Shine}, Richard A. and {Slater}, Gregory L. and {Tarbell}, Theodore D. and {Title}, Alan M. and {Okamoto}, Takenori J. and {Ichimoto}, Kiyoshi and {Katsukawa}, Yukio and {Suematsu}, Yoshinori and {Tsuneta}, Saku and {Lites}, Bruce W. and {Shimizu}, Toshifumi},
        title = "{Hinode SOT Observations of Solar Quiescent Prominence Dynamics}",
      journal = {\apjl},
         year = 2008,
        month = mar,
       volume = {676},
       number = {1},
        pages = {L89},
          doi = {10.1086/587171},
       adsurl = {https://ui.adsabs.harvard.edu/abs/2008ApJ...676L..89B}
}

@ARTICLE{2010ApJ...716.1288B,
       author = {{Berger}, Thomas E. and {Slater}, Gregory and {Hurlburt}, Neal and {Shine}, Richard and {Tarbell}, Theodore and {Title}, Alan and {Lites}, Bruce W. and {Okamoto}, Takenori J. and {Ichimoto}, Kiyoshi and {Katsukawa}, Yukio and {Magara}, Tetsuya and {Suematsu}, Yoshinori and {Shimizu}, Toshifumi},
        title = "{Quiescent Prominence Dynamics Observed with the Hinode Solar Optical Telescope. I. Turbulent Upflow Plumes}",
      journal = {\apj},
         year = 2010,
        month = jun,
       volume = {716},
       number = {2},
        pages = {1288-1307},
          doi = {10.1088/0004-637X/716/2/1288},
       adsurl = {https://ui.adsabs.harvard.edu/abs/2010ApJ...716.1288B}
}

@ARTICLE{2017ApJ...850...60B,
       author = {{Berger}, Thomas and {Hillier}, Andrew and {Liu}, Wei},
        title = "{Quiescent Prominence Dynamics Observed with the Hinode Solar Optical Telescope. II. Prominence Bubble Boundary Layer Characteristics and the Onset of a Coupled Kelvin-Helmholtz Rayleigh-Taylor Instability}",
      journal = {\apj},
         year = 2017,
        month = nov,
       volume = {850},
       number = {1},
          eid = {60},
        pages = {60},
          doi = {10.3847/1538-4357/aa95b6},
archivePrefix = {arXiv},
       eprint = {1707.05265},
 primaryClass = {astro-ph.SR},
       adsurl = {https://ui.adsabs.harvard.edu/abs/2017ApJ...850...60B}
}

@ARTICLE{2011Natur.472..197B,
       author = {{Berger}, Thomas and {Testa}, Paola and {Hillier}, Andrew and {Boerner}, Paul and {Low}, Boon Chye and {Shibata}, Kazunari and {Schrijver}, Carolus and {Tarbell}, Ted and {Title}, Alan},
        title = "{Magneto-thermal convection in solar prominences}",
      journal = {\nat},
         year = 2011,
        month = apr,
       volume = {472},
       number = {7342},
        pages = {197-200},
          doi = {10.1038/nature09925},
       adsurl = {https://ui.adsabs.harvard.edu/abs/2011Natur.472..197B}
}

@ARTICLE{2021ApJ...923L..10C,
       author = {{Chen}, Changxue and {Su}, Yang and {Xue}, Jianchao and {Gan}, Weiqun and {Huang}, Yu},
        title = "{Solar Prominence Bubble and Plumes Caused By an Eruptive Magnetic Flux Rope}",
      journal = {\apjl},
         year = 2021,
        month = dec,
       volume = {923},
       number = {1},
          eid = {L10},
        pages = {L10},
          doi = {10.3847/2041-8213/ac3bd0},
       adsurl = {https://ui.adsabs.harvard.edu/abs/2021ApJ...923L..10C}
}

@ARTICLE{2026SCPMA..6949611C,
       author = {{Chen}, Huanxin and {Xia}, Chun and {Chen}, Hechao},
        title = "{The spontaneous genesis of solar prominence structures driven by supergranulation in three-dimensional simulations}",
      journal = {Science China Physics, Mechanics, and Astronomy},
         year = 2026,
        month = feb,
       volume = {69},
       number = {4},
          eid = {249611},
        pages = {249611},
          doi = {10.1007/s11433-025-2858-5},
archivePrefix = {arXiv},
       eprint = {2511.13252},
 primaryClass = {astro-ph.SR},
       adsurl = {https://ui.adsabs.harvard.edu/abs/2026SCPMA..6949611C}
}

@ARTICLE{2020RAA....20..166C,
       author = {{Chen}, Peng-Fei and {Xu}, Ao-Ao and {Ding}, Ming-De},
        title = "{Some interesting topics provoked by the solar filament research in the past decade}",
      journal = {Research in Astronomy and Astrophysics},
         year = 2020,
        month = oct,
       volume = {20},
       number = {10},
          eid = {166},
        pages = {166},
          doi = {10.1088/1674-4527/20/10/166},
archivePrefix = {arXiv},
       eprint = {2010.02462},
 primaryClass = {astro-ph.SR},
       adsurl = {https://ui.adsabs.harvard.edu/abs/2020RAA....20..166C}
}

@ARTICLE{2012ApJ...761....9D,
       author = {{Dud{\'\i}k}, J. and {Aulanier}, G. and {Schmieder}, B. and {Zapi{\'o}r}, M. and {Heinzel}, P.},
        title = "{Magnetic Topology of Bubbles in Quiescent Prominences}",
      journal = {\apj},
         year = 2012,
        month = dec,
       volume = {761},
       number = {1},
          eid = {9},
        pages = {9},
          doi = {10.1088/0004-637X/761/1/9},
       adsurl = {https://ui.adsabs.harvard.edu/abs/2012ApJ...761....9D}
}

@ARTICLE{2007ITIP...16.2080D,
       author = {{Dabov}, Kostadin and {Foi}, Alessandro and {Katkovnik}, Vladimir and {Egiazarian}, Karen},
        title = "{Image Denoising by Sparse 3-D Transform-Domain Collaborative Filtering}",
      journal = {IEEE Transactions on Image Processing},
         year = 2007,
        month = jan,
       volume = {16},
       number = {8},
        pages = {2080-2095},
          doi = {10.1109/TIP.2007.901238},
       adsurl = {https://ui.adsabs.harvard.edu/abs/2007ITIP...16.2080D}
}

@ARTICLE{2021ApJ...911L...9G,
       author = {{Guo}, Yilin and {Hou}, Yijun and {Li}, Ting and {Zhang}, Jun},
        title = "{Reconstructing 3D Magnetic Topology of On-disk Prominence Bubbles from Stereoscopic Observations}",
      journal = {\apjl},
         year = 2021,
        month = apr,
       volume = {911},
       number = {1},
          eid = {L9},
        pages = {L9},
          doi = {10.3847/2041-8213/abee92},
archivePrefix = {arXiv},
       eprint = {2103.07860},
 primaryClass = {astro-ph.SR},
       adsurl = {https://ui.adsabs.harvard.edu/abs/2021ApJ...911L...9G}
}

@ARTICLE{2024ApJ...970..110G,
       author = {{Guo}, Yilin and {Hou}, Yijun and {Li}, Ting and {Shen}, Yuandeng and {Wang}, Jincheng and {Zhang}, Jun and {Zheng}, Jianchuan and {Wang}, Dong and {Mei}, Lin},
        title = "{Formation and Evolution of Transient Prominence Bubbles Driven by Erupting Minifilaments}",
      journal = {\apj},
         year = 2024,
        month = aug,
       volume = {970},
       number = {2},
          eid = {110},
        pages = {110},
          doi = {10.3847/1538-4357/ad54b8},
archivePrefix = {arXiv},
       eprint = {2405.04725},
 primaryClass = {astro-ph.SR},
       adsurl = {https://ui.adsabs.harvard.edu/abs/2024ApJ...970..110G}
}

@ARTICLE{2014A&A...567A.123G,
       author = {{Gun{\'a}r}, S. and {Schwartz}, P. and {Dud{\'\i}k}, J. and {Schmieder}, B. and {Heinzel}, P. and {Jur{\v{c}}{\'a}k}, J.},
        title = "{Magnetic field and radiative transfer modelling of a quiescent prominence}",
      journal = {\aap},
         year = 2014,
        month = jul,
       volume = {567},
          eid = {A123},
        pages = {A123},
          doi = {10.1051/0004-6361/201322777},
       adsurl = {https://ui.adsabs.harvard.edu/abs/2014A&A...567A.123G}
}

@ARTICLE{2023ApJ...959...69H,
       author = {{Hou}, Yijun and {Li}, Chuan and {Li}, Ting and {Su}, Jiangtao and {Qiu}, Ye and {Yang}, Shuhong and {Yang}, Liheng and {Li}, Leping and {Guo}, Yilin and {Hou}, Zhengyong and {Song}, Qiao and {Bai}, Xianyong and {Zhou}, Guiping and {Ding}, Mingde and {Gan}, Weiqun and {Deng}, Yuanyong},
        title = "{Partial Eruption of Solar Filaments. I. Configuration and Formation of Double-decker Filaments}",
      journal = {\apj},
         year = 2023,
        month = dec,
       volume = {959},
       number = {2},
          eid = {69},
        pages = {69},
          doi = {10.3847/1538-4357/ad08bd},
archivePrefix = {arXiv},
       eprint = {2311.00456},
 primaryClass = {astro-ph.SR},
       adsurl = {https://ui.adsabs.harvard.edu/abs/2023ApJ...959...69H}
}

@ARTICLE{2020A&A...640A.101H,
       author = {{Hou}, Y.~J. and {Li}, T. and {Song}, Z.~P. and {Zhang}, J.},
        title = "{External reconnection and resultant reconfiguration of overlying magnetic fields during sympathetic eruptions of two filaments}",
      journal = {\aap},
         year = 2020,
        month = aug,
       volume = {640},
          eid = {A101},
        pages = {A101},
          doi = {10.1051/0004-6361/202038348},
archivePrefix = {arXiv},
       eprint = {2006.06191},
 primaryClass = {astro-ph.SR},
       adsurl = {https://ui.adsabs.harvard.edu/abs/2020A&A...640A.101H}
}

@ARTICLE{2012ApJ...746..120H,
       author = {{Hillier}, Andrew and {Berger}, Thomas and {Isobe}, Hiroaki and {Shibata}, Kazunari},
        title = "{Numerical Simulations of the Magnetic Rayleigh-Taylor Instability in the Kippenhahn-Schl{\"u}ter Prominence Model. I. Formation of Upflows}",
      journal = {\apj},
         year = 2012,
        month = feb,
       volume = {746},
       number = {2},
          eid = {120},
        pages = {120},
          doi = {10.1088/0004-637X/746/2/120},
       adsurl = {https://ui.adsabs.harvard.edu/abs/2012ApJ...746..120H}
}

@ARTICLE{2012ApJ...756..110H,
       author = {{Hillier}, Andrew and {Isobe}, Hiroaki and {Shibata}, Kazunari and {Berger}, Thomas},
        title = "{Numerical Simulations of the Magnetic Rayleigh-Taylor Instability in the Kippenhahn-Schl{\"u}ter Prominence Model. II. Reconnection-triggered Downflows}",
      journal = {\apj},
         year = 2012,
        month = sep,
       volume = {756},
       number = {2},
          eid = {110},
        pages = {110},
          doi = {10.1088/0004-637X/756/2/110},
archivePrefix = {arXiv},
       eprint = {1106.2613},
 primaryClass = {astro-ph.SR},
       adsurl = {https://ui.adsabs.harvard.edu/abs/2012ApJ...756..110H}
}

@ARTICLE{2018RvMPP...2....1H,
       author = {{Hillier}, Andrew},
        title = "{The magnetic Rayleigh-Taylor instability in solar prominences}",
      journal = {Reviews of Modern Plasma Physics},
         year = 2018,
        month = dec,
       volume = {2},
       number = {1},
          eid = {1},
        pages = {1},
          doi = {10.1007/s41614-017-0013-2},
       adsurl = {https://ui.adsabs.harvard.edu/abs/2018RvMPP...2....1H}
}

@ARTICLE{2022SCPMA..6539611L,
       author = {{Li}, Dong and {Xue}, Jianchao and {Yuan}, Ding and {Ning}, Zongjun},
        title = "{Persistent fast kink magnetohydrodynamic waves detected in a quiescent prominence}",
      journal = {Science China Physics, Mechanics, and Astronomy},
         year = 2022,
        month = mar,
       volume = {65},
       number = {3},
          eid = {239611},
        pages = {239611},
          doi = {10.1007/s11433-021-1836-y},
archivePrefix = {arXiv},
       eprint = {2201.07535},
 primaryClass = {astro-ph.SR},
       adsurl = {https://ui.adsabs.harvard.edu/abs/2022SCPMA..6539611L}
}

@ARTICLE{2014RAA....14..705L,
       author = {{Liu}, Zhong and {Xu}, Jun and {Gu}, Bo-Zhong and {Wang}, Sen and {You}, Jian-Qi and {Shen}, Long-Xiang and {Lu}, Ru-Wei and {Jin}, Zhen-Yu and {Chen}, Lin-Fei and {Lou}, Ke and {Li}, Zhi and {Liu}, Guang-Qian and {Xu}, Zhi and {Rao}, Chang-Hui and {Hu}, Qi-Qian and {Li}, Ru-Feng and {Fu}, Hao-Wen and {Wang}, Feng and {Bao}, Men-Xian and {Wu}, Ming-Chan and {Zhang}, Bo-Rong},
        title = "{New vacuum solar telescope and observations with high resolution}",
      journal = {Research in Astronomy and Astrophysics},
         year = 2014,
        month = jun,
       volume = {14},
       number = {6},
          eid = {705-718},
        pages = {705-718},
          doi = {10.1088/1674-4527/14/6/009},
archivePrefix = {arXiv},
       eprint = {1403.6896},
 primaryClass = {astro-ph.IM},
       adsurl = {https://ui.adsabs.harvard.edu/abs/2014RAA....14..705L}
}

@ARTICLE{2010SSRv..151..243L,
       author = {{Labrosse}, N. and {Heinzel}, P. and {Vial}, J.-C. and {Kucera}, T. and {Parenti}, S. and {Gun{\'a}r}, S. and {Schmieder}, B. and {Kilper}, G.},
        title = "{Physics of Solar Prominences: I{\textemdash}Spectral Diagnostics and Non-LTE Modelling}",
      journal = {\ssr},
         year = 2010,
        month = apr,
       volume = {151},
       number = {4},
        pages = {243-332},
          doi = {10.1007/s11214-010-9630-6},
archivePrefix = {arXiv},
       eprint = {1001.1620},
 primaryClass = {astro-ph.SR},
       adsurl = {https://ui.adsabs.harvard.edu/abs/2010SSRv..151..243L}
}

@ARTICLE{2018ApJ...863..192L,
       author = {{Li}, Dong and {Shen}, Yuandeng and {Ning}, Zongjun and {Zhang}, Qingmin and {Zhou}, Tuanhui},
        title = "{Two Kinds of Dynamic Behavior in a Quiescent Prominence Observed by the NVST}",
      journal = {\apj},
         year = 2018,
        month = aug,
       volume = {863},
       number = {2},
          eid = {192},
        pages = {192},
          doi = {10.3847/1538-4357/aad33f},
archivePrefix = {arXiv},
       eprint = {1807.03942},
 primaryClass = {astro-ph.SR},
       adsurl = {https://ui.adsabs.harvard.edu/abs/2018ApJ...863..192L}
}

@ARTICLE{2015SoPh..290.1703M,
       author = {{McCauley}, P.~I. and {Su}, Y.~N. and {Schanche}, N. and {Evans}, K.~E. and {Su}, C. and {McKillop}, S. and {Reeves}, K.~K.},
        title = "{Prominence and Filament Eruptions Observed by the Solar Dynamics Observatory: Statistical Properties, Kinematics, and Online Catalog}",
      journal = {\solphys},
         year = 2015,
        month = jun,
       volume = {290},
       number = {6},
        pages = {1703-1740},
          doi = {10.1007/s11207-015-0699-7},
archivePrefix = {arXiv},
       eprint = {1505.02090},
 primaryClass = {astro-ph.SR},
       adsurl = {https://ui.adsabs.harvard.edu/abs/2015SoPh..290.1703M}
}

@ARTICLE{1980RSPSB.207..187M,
       author = {{Marr}, D. and {Hildreth}, E.},
        title = "{Theory of Edge Detection}",
      journal = {Proceedings of the Royal Society of London Series B},
         year = 1980,
        month = feb,
       volume = {207},
       number = {1167},
        pages = {187-217},
          doi = {10.1098/rspb.1980.0020},
       adsurl = {https://ui.adsabs.harvard.edu/abs/1980RSPSB.207..187M}
}

@ARTICLE{2014LRSP...11....1P,
       author = {{Parenti}, Susanna},
        title = "{Solar Prominences: Observations}",
      journal = {Living Reviews in Solar Physics},
         year = 2014,
        month = dec,
       volume = {11},
       number = {1},
          eid = {1},
        pages = {1},
          doi = {10.12942/lrsp-2014-1},
       adsurl = {https://ui.adsabs.harvard.edu/abs/2014LRSP...11....1P}
}

@ARTICLE{2010SoPh..267...75R,
       author = {{Ryutova}, M. and {Berger}, T. and {Frank}, Z. and {Tarbell}, T. and {Title}, A.},
        title = "{Observation of Plasma Instabilities in Quiescent Prominences}",
      journal = {\solphys},
         year = 2010,
        month = nov,
       volume = {267},
       number = {1},
        pages = {75-94},
          doi = {10.1007/s11207-010-9638-9},
       adsurl = {https://ui.adsabs.harvard.edu/abs/2010SoPh..267...75R}
}

@ARTICLE{2014ApJ...795..130S,
       author = {{Shen}, Yuandeng and {Liu}, Ying D. and {Chen}, P.~F. and {Ichimoto}, Kiyoshi},
        title = "{Simultaneous Transverse Oscillations of a Prominence and a Filament and Longitudinal Oscillation of Another Filament Induced by a Single Shock Wave}",
      journal = {\apj},
         year = 2014,
        month = nov,
       volume = {795},
       number = {2},
          eid = {130},
        pages = {130},
          doi = {10.1088/0004-637X/795/2/130},
archivePrefix = {arXiv},
       eprint = {1409.1304},
 primaryClass = {astro-ph.SR},
       adsurl = {https://ui.adsabs.harvard.edu/abs/2014ApJ...795..130S}
}

@ARTICLE{2015ApJ...814L..17S,
       author = {{Shen}, Yuandeng and {Liu}, Yu and {Liu}, Ying D. and {Chen}, P.~F. and {Su}, Jiangtao and {Xu}, Zhi and {Liu}, Zhong},
        title = "{Fine Magnetic Structure and Origin of Counter-streaming Mass Flows in a Quiescent Solar Prominence}",
      journal = {\apjl},
         year = 2015,
        month = nov,
       volume = {814},
       number = {1},
          eid = {L17},
        pages = {L17},
          doi = {10.1088/2041-8205/814/1/L17},
archivePrefix = {arXiv},
       eprint = {1511.02489},
 primaryClass = {astro-ph.SR},
       adsurl = {https://ui.adsabs.harvard.edu/abs/2015ApJ...814L..17S}
}

@ARTICLE{2025ApJ...990L..64W,
       author = {{Wang}, Wensi and {Liu}, Rui and {Luo}, Runbin and {Yan}, Xiaoli},
        title = "{High-resolution Observation of Solar Prominence Plumes Induced by Enhanced Spicular Activity}",
      journal = {\apjl},
         year = 2025,
        month = sep,
       volume = {990},
       number = {2},
          eid = {L64},
        pages = {L64},
          doi = {10.3847/2041-8213/adfecf},
archivePrefix = {arXiv},
       eprint = {2508.13456},
 primaryClass = {astro-ph.SR},
       adsurl = {https://ui.adsabs.harvard.edu/abs/2025ApJ...990L..64W}
}

@ARTICLE{2022A&A...659A..76W,
       author = {{Wang}, Jincheng and {Yan}, Xiaoli and {Xue}, Zhike and {Yang}, Liheng and {Li}, Qiaoling and {Chen}, Hechao and {Xia}, Chun and {Liu}, Zhong},
        title = "{A formation mechanism for the large plumes in the prominence}",
      journal = {\aap},
         year = 2022,
        month = mar,
       volume = {659},
          eid = {A76},
        pages = {A76},
          doi = {10.1051/0004-6361/202142584},
archivePrefix = {arXiv},
       eprint = {2202.08521},
 primaryClass = {astro-ph.SR},
       adsurl = {https://ui.adsabs.harvard.edu/abs/2022A&A...659A..76W}
}

@ARTICLE{2024ApJ...965L..28W,
       author = {{Wang}, Jincheng and {Li}, Dong and {Li}, Chuan and {Hou}, Yijun and {Xue}, Zhike and {Xu}, Zhe and {Yang}, Liheng and {Li}, Qiaoling},
        title = "{Negative-energy Waves in the Vertical Threads of a Solar Prominence}",
      journal = {\apjl},
         year = 2024,
        month = apr,
       volume = {965},
       number = {2},
          eid = {L28},
        pages = {L28},
          doi = {10.3847/2041-8213/ad3af8},
archivePrefix = {arXiv},
       eprint = {2404.03199},
 primaryClass = {astro-ph.SR},
       adsurl = {https://ui.adsabs.harvard.edu/abs/2024ApJ...965L..28W}
}

@ARTICLE{2021RAA....21..222X,
       author = {{Xue}, Jian-Chao and {Vial}, Jean-Claude and {Su}, Yang and {Li}, Hui and {Xu}, Zhi and {Su}, Ying-Na and {Zhou}, Tuan-Hui and {Li}, Zhen-Tong},
        title = "{High-resolution observations of prominence plume formation with the new vacuum solar telescope}",
      journal = {Research in Astronomy and Astrophysics},
         year = 2021,
        month = nov,
       volume = {21},
       number = {9},
          eid = {222},
        pages = {222},
          doi = {10.1088/1674-4527/21/9/222},
       adsurl = {https://ui.adsabs.harvard.edu/abs/2021RAA....21..222X}
}

@ARTICLE{2020ScChE..63.1656Y,
       author = {{Yan}, XiaoLi and {Liu}, Zhong and {Zhang}, Jun and {Xu}, Zhi},
        title = "{Research progress based on observations of the New Vacuum Solar Telescope}",
      journal = {Science in China E: Technological Sciences},
         year = 2020,
        month = sep,
       volume = {63},
       number = {9},
        pages = {1656-1674},
          doi = {10.1007/s11431-019-1463-6},
archivePrefix = {arXiv},
       eprint = {1910.09127},
 primaryClass = {astro-ph.SR},
       adsurl = {https://ui.adsabs.harvard.edu/abs/2020ScChE..63.1656Y}
}

@ARTICLE{2025ApJ...981..139Y,
       author = {{Yan}, Xiaoli and {Xue}, Zhike and {Wang}, Jincheng and {Chen}, Pengfei and {Ji}, Kaifan and {Xia}, Chun and {Yang}, Liheng and {Kong}, Defang and {Xu}, Zhe and {Zhou}, Yian and {Li}, Qiaoling},
        title = "{Simultaneous Existence of Oscillations, Counterstreaming Flows, and Mass Injections in Solar Quiescent Prominences}",
      journal = {\apj},
         year = 2025,
        month = mar,
       volume = {981},
       number = {2},
          eid = {139},
        pages = {139},
          doi = {10.3847/1538-4357/adb39e},
archivePrefix = {arXiv},
       eprint = {2502.04114},
 primaryClass = {astro-ph.SR},
       adsurl = {https://ui.adsabs.harvard.edu/abs/2025ApJ...981..139Y}
}

@ARTICLE{2019ApJ...884L..51Y,
       author = {{Yuan}, Ding and {Shen}, Yuandeng and {Liu}, Yu and {Li}, Hongbo and {Feng}, Xueshang and {Keppens}, Rony},
        title = "{Multilayered Kelvin-Helmholtz Instability in the Solar Corona}",
      journal = {\apjl},
         year = 2019,
        month = oct,
       volume = {884},
       number = {2},
          eid = {L51},
        pages = {L51},
          doi = {10.3847/2041-8213/ab4bcd},
archivePrefix = {arXiv},
       eprint = {1910.05710},
 primaryClass = {astro-ph.SR},
       adsurl = {https://ui.adsabs.harvard.edu/abs/2019ApJ...884L..51Y}
}

@ARTICLE{2012A&A...542A..52Z,
       author = {{Zhang}, Q.~M. and {Chen}, P.~F. and {Xia}, C. and {Keppens}, R.},
        title = "{Observations and simulations of longitudinal oscillations of an active region prominence}",
      journal = {\aap},
         year = 2012,
        month = jun,
       volume = {542},
          eid = {A52},
        pages = {A52},
          doi = {10.1051/0004-6361/201218786},
archivePrefix = {arXiv},
       eprint = {1204.3787},
 primaryClass = {astro-ph.SR},
       adsurl = {https://ui.adsabs.harvard.edu/abs/2012A&A...542A..52Z}
}
\bibliographystyle{aasjournalv7}

\end{document}